\documentclass[sigconf]{acmart}

\setcopyright{none} %

\setcopyright{none}
\renewcommand\footnotetextcopyrightpermission[1]{} 
\usepackage{listings}                       %

\usepackage{multirow}
\usepackage{graphicx}
\usepackage{paralist}
\usepackage[lined,boxed]{algorithm2e}

\usepackage{textcomp}
\usepackage{xcolor}
\usepackage{flushend}
\usepackage{booktabs}
\usepackage{framed}
\usepackage{enumerate}
\usepackage{graphicx,subfig}
\usepackage{xspace}
\usepackage{boldline}

\usepackage[marginal,hang,stable]{footmisc}
\renewcommand{\footnoterule}{%
  \kern -3pt
  \hrule width 1in 
  \kern 2pt
}

\newcommand{\systemName}{\textsc{Cerberus}\xspace}

\newif\ifcomment
\commenttrue

\ifcomment
\newcommand\edc[1]{\textbf{\textcolor{blue}{EDC: #1}}	}
\newcommand\yun[1]{\textbf{\textcolor{brown}{Yun: #1}}	}
\newcommand\mn[1]{\textbf{\textcolor{orange}{Mohammad: #1}}	}
\newcommand\yf[1]{\textbf{\textcolor{cyan}{Yufei: #1}}	}
\newcommand\gs[1]{\textbf{\textcolor{red}{Gianluca: #1}}	}

\else
\newcommand\edc[1]{}
\newcommand\mn[1]{}
\newcommand\yf[1]{}
\newcommand\gs[1]{}
\newcommand\yun[1]{}

\fi

\newcommand{\descr}[1]{\smallskip\noindent\textbf{#1}}

\newtheorem{ldp-definition}{Definition}
\newtheorem{dp-definition}{Definition}
\newcommand\numberoforgs{\mbox{5,419}}
\newcommand\totalnumberofmachines{\mbox{2,001,746}}
\newcommand\numberofeventstypes{\mbox{1,465}}
\newcommand\totalnumberofevents{\mbox{34,846,425}}

\begin{document}
\title{\systemName: Exploring Federated Prediction of Security Events}\titlenote{\bf\em Published in ACM CCS 2022. Please cite the CCS version.}
\author{Mohammad Naseri}
\affiliation{%
  \institution{University College London}
  \country{}
}
\email{mohammad.naseri.19@ucl.ac.uk}

\author{Yufei Han}
\affiliation{%
  \institution{Inria Rennes}
  \country{}
}
\email{yufei.han@inria.fr}

\author{Enrico Mariconti}
\affiliation{%
  \institution{University College London}
  \country{}
}
\email{e.mariconti@ucl.ac.uk}

\author{Yun Shen}
\authornote{Work partially done while the author was with NortonLifeLock.}
\affiliation{%
  \institution{NetApp}
  \country{}
}
\email{Yun.Shen@netapp.com}

\author{Gianluca Stringhini}
\affiliation{%
  \institution{Boston University}
  \country{}
}
\email{gian@bu.edu}

\author{Emiliano De Cristofaro}
\affiliation{%
  \institution{University College London}
  \country{}
}
\email{e.decristofaro@ucl.ac.uk}

\begin{abstract}
Modern defenses against cyberattacks increasingly rely on proactive approaches, e.g., to predict the adversary's next actions based on past events. 
Building accurate prediction models requires knowledge from many organizations; alas, this entails disclosing sensitive information, such as network structures, security postures, and policies, which might often be undesirable or outright impossible.

In this paper, we explore the feasibility of using Federated Learning (FL) to predict future security events.
To this end, we introduce \systemName, a system enabling collaborative training of Recurrent Neural Network (RNN) models for participating organizations.
The intuition is that FL could potentially offer a middle-ground between the non-private approach where the training data is pooled at a central server and the low-utility alternative of only training local models.
We instantiate \systemName on a dataset obtained from a major security company's intrusion prevention product and evaluate it vis-\`a-vis utility, robustness, and privacy, as well as how participants contribute to and benefit from the system.
Overall, our work sheds light on both the positive aspects and the challenges of using FL for this task and paves the way for deploying federated approaches to predictive security.
\end{abstract}

\maketitle
\pagestyle{plain}

\section{Introduction}
\label{Introduction}

Modern security breaches often happen in multiple phases, with attackers progressively gaining a more significant foothold in an organization.
Alas, attacks are often detected at later stages or even after they have been completed, making remediation much more difficult. %
As a result, predicting an attack early on can give organizations a significant advantage, enabling them to take proactive rather than reactive countermeasures. 

Prior work introduced systems to predict security events~\cite{shen2018tiresias,soska2014automatically,bilge2017riskteller,liu2015cloudy,liu2015predicting,sabottke2015vulnerability}: the general approach is to learn from history, characterizing previous attack events and using this knowledge to predict future ones.
However, these systems typically require collecting events from organizations and storing them in a centralized platform to train a ML-driven prediction model. 
These records often include privacy-sensitive metadata, including machine ID, event description, timestamp, action taken, etc.
Moreover, disclosing security events can reveal sensitive information about network structures, security policies, security postures, etc. %

A possible alternative solution would be to adopt a local-only procedure where each organization only trains a prediction system on their own data. 
However, %
one would not have access to the intelligence and knowledge available from a (more) global view of security events; put simply, it would be significantly harder to learn about emerging attacks that target other organizations.

Overall, in many scenarios, confidentiality concerns would, in effect, make it impossible to perform security event prediction as disclosing data is not possible, and local-only training is ineffective.
As a result, we investigate the feasibility of using collaborative learning to benefit from participating organizations' knowledge without requiring data disclosure. 
In particular, we turn to Federated Learning (FL)~\cite{mcmahan2017communication}, a popular technique for training machine learning models collaboratively %
based on aggregated model updates.

Ostensibly, this raises several research challenges.
First, it is unclear what the resulting accuracy of this approach would be or whether organizations would benefit from participating in the system. 
Moreover, even though with FL raw data never leaves the ``premises,'' prior work shows that FL  does suffer from privacy and robustness vulnerabilities~\cite{melis2019exploiting,zhu2019deep,nasr2019comprehensive,bagdasaryan2020backdoor}.
This prompts the need for a thorough experimental methodology, taking different data distributions and settings into account to evaluate utility and security in real-world settings. %

\descr{Research Questions \& Roadmap.} In a nutshell, our work identifies and aims to address three main research questions: 
\begin{enumerate}
	\item Is it feasible to build an FL-based system to predict security events?
	
	\item How can we meaningfully analyze the utility of such a system? How would the data distribution across different participants affect the accuracy of the prediction model? How do different participants contribute to or benefit from the federation?
	
	\item How vulnerable is FL-based security event prediction to robustness and privacy attacks, and do available defenses mitigate them effectively?

\end{enumerate}

To answer these questions, we design \systemName, an FL system for predicting security events (Section~\ref{sec:flachitecture}). 
\systemName uses a Recurrent Neural Network (RNN) to train a model learning from the history of security events and predicting future security incidents. %
The system does not collect security events from organizations; instead, each participant obtains a model to be trained on their dataset. 

We evaluate \systemName on an intrusion prevention product run by a major security company, using a dataset including nearly 5K organizations and 35M security events (see Section~\ref{sec:dataset}). 
More precisely, we analyze \systemName along three axes: \textit{utility}, \textit{robustness}, and \textit{privacy} (see Section~\ref{sec:utility},~\ref{sec:robustness}, and~\ref{sec:privacy}, respectively). %

\descr{Methodology.} Our experimental evaluations are conducted over different data distributions.
We define a metric to measure how Non-Independent and Non-Identical the data distributions of different participants are (the {\em Non-IIDness score}) and synthesize different distribution settings based on this score. %
For instance, we experiment with a setting involving so-called {\em knowledgeable participants}, i.e., organizations with data instances from {\em all} classes.  

We then evaluate how much and how many different participants contribute to the aggregated model's accuracy. 
To do so, we remove a participant from the system and estimate the impact on the aggregated model's precision; %
we call this metric the {\em contribution impact}.
However, it would be prohibitively expensive to do so for all participants because of the computational overhead of FL's re-training phase.
Thus, we use the {\em influence score}, as defined in~\cite{koh2017understanding}, to measure the impact of each participant's training dataset on its local model.
We then compute the contribution impact for the participants with the highest influence score and show that the latter metric can be used as a proxy for the former to shed light on what data makes the federated model work well.
We also analyze the benefits of participating in \systemName by comparing the utility of the global model vs.~the local model. %
We show that, for the participants with the highest contribution, their ``gain'' from federating is dependent on the data distribution.

Finally, we analyze \systemName's vulnerability to robustness and privacy attacks, i.e., measuring the effectiveness of poisoned data contributed by adversarial participants and the privacy leakage from the model updates.
\descr{Main Findings.} Overall, we find that: 
\begin{itemize}
	\item[$\bullet$] The utility of the FL-based system, \systemName, is relatively lower than in the centralized alternative, where the server gathers raw data from all participants and trains the model.
	For instance, the F1-score decreases from 0.83 to 0.70, accuracy from 0.85 to 0.78 using what we define as the primary distribution. Precision, computed w.r.t.~1,465 possible security event types, goes from 0.84 to 0.69.
	The accuracy drop is, essentially, the ``cost'' of doing away with the central server gathering security events from all participants.
\item[$\bullet$] In extreme non-IID distributions, FL accuracy drops further. 
For instance, the F1-score goes down to 0.65 in the extreme (artificial) setting where each participant is assigned one class, suggesting that FL-based approaches might not always be feasible if the security events are distributed this way across organizations.

\item[$\bullet$] The contribution impact is a correct metric to measure participants' contribution to the FL aggregated model, as knowledgeable participants, i.e., those with instances from all classes, have the highest value. 
Some organizations contribute significantly to the performance of \systemName but do not benefit much as their local datasets are already ``rich'' with instances from all the classes.
In line with other work on FL~\cite{yu2020salvaging}, this suggests that the issues of benefit-vs-contribution and how to incentivize participation need to be taken into account.

	\item[$\bullet$] Distributed backdoor poisoning attacks are relatively effective at undermining robustness while decreasing the main task precision by a negligible amount. 
	For instance, with 1\% of participants being controlled by an adversary, the backdoor attack reaches an accuracy of 0.94, while the main task's precision decreases from 0.69 to 0.65. 
However, defenses like norm bounding~\cite{sun2019can}, Weak Differential Privacy (DP)~\cite{sun2019can}, and Centralized Differential Privacy (CDP)~\cite{mcmahan2017learning,geyer2017differentially,DBLP:journals/corr/abs-2009-03561} are quite effective across the board---as opposed to \textit{Trimmed Mean}~\cite{yin2018byzantine} and \textit{Krum}~\cite{blanchard2017machine}, which do not work in Non-IID data distributions.
	
	\item[$\bullet$] Membership inference attacks~\cite{nasr2019comprehensive,zhang2020gan} are effective, but only with a few participants.
	In these settings, the best potential defense, CDP, is hard to deploy as the model ends up not converging due to the noise needed to be added to the model updates. 
\end{itemize}

\descr{Contributions.} The main contribution of our work is to explore the feasibility of using FL to collaboratively train RNNs and predict security events.
We do so in four main steps:
1) we introduce \systemName, a (generic) system using FL and RNN to collaboratively predict security events;
2) we define appropriate metrics to measure the contribution of entities to the system, as well as the benefits for them;
3) we evaluate the vulnerability to distributed backdoor data poisoning~\cite{bagdasaryan2020backdoor} as well as privacy-leakage attacks~\cite{nasr2019comprehensive,zhang2020gan}, and the effectiveness of state-of-the-art defenses; and
4) we discuss open challenges and a roadmap for future work in this space.

\section{Federated Learning Background}\label{preliminary}
We now review the notion of Federated Learning (FL), as well as the instantiation we use ($FedAvg$~\cite{mcmahan2017communication}). 
We also present attacks against FL and available mitigations.
Readers familiar with these notions can skip this section without loss of continuity.

\subsection{FedAvg}\label{preliminary:fl}
Federated Learning (FL) is a distributed learning setting used to collaboratively train models with multiple participants~\cite{mcmahan2017communication}.
Unlike traditional centralized approaches, training data instances are not pooled at a central server.
Each participant trains their own model locally, on their datasets, and shares updated parameters with a server, which aggregates the parameters and returns the result to the participants.
Typically, this happens over multiple rounds; eventually, the model converges and the parameters are finalized. 

In this paper, we consider the FL instantiation presented in~\cite{mcmahan2017communication}, which relies on the $FedAvg$ (Federated Averaging) algorithm. %
The model is training iteratively; let $\theta^{t}_{global}$ denote the latest global model aggregated by the central server at the iteration step $t$, and $m_{i}$ (for $i=1,2,3,4,\ldots,K$) denote the devices of all $K$ participants. %
The central server first broadcasts $\theta^{t}_{global}$ to all $m_{i}$, then, every device (say the $k$-th) initializes $\theta^{t}_{i}$ as $\theta^{t}_{i}= \theta^{t}_{global}$. 
The $i$-th participant performs $E$ (where $E\geq{1}$) local updates:  	\vspace{-0.05cm}
\begin{equation}\small
	\theta^{t}_{i}(k+1) = \theta^{t}_{i}(k) - \eta_{k}\nabla \ell(\theta^{t}_{i}(k),\{x_{i,j},y_{i,j}\})  	\vspace{-0.05cm}
\end{equation}
where $k=1,2,3,...,E$ denotes the local training step.
Note that $\ell$ is the classification loss function used by the $i$-th participants, while $\{x_{i,j},y_{i,j}\}$  (where $j=1,2,3,\ldots,|V_{i}|$) are the training instances hosted locally by the $i$-th participant. 
$|V_{i}|$ is the total number of training instances on the i-th participant,
Finally, $\theta^{t}_{i}(k)$ is the local model updated at the $k$-th step of the stochastic gradient descent performed locally; note that $\theta^{t}_{i}(k)$ = $\theta^{t}_{i}$ = $\theta^{t}_{global}$ when $k=0$.  

After the participants finish training their local models, they submit them to the central server. 
The global model derived at the central server is then aggregated by averaging the local models:\vspace{-0.1cm}
\begin{equation}\small
	\theta^{t+1}_{global} = \sum_{i=1}^{K} \theta^{t}_{i}(E)\vspace{-0.1cm}
\end{equation}
Note that $\theta^{t}_{i}(E)$ denotes the models trained locally, after $E$ rounds of gradient descent. 

More advanced averaging techniques may be applied to \textit{FedAvg}, e.g., taking the weighted average of local models to derive the global model \cite{LiHYWZ20}. 
These techniques may address settings involving heterogeneous data across participants.
However, compared to the standard \textit{FedAvg}, they only slightly improve the training convergence in the not independent and identically distributed scenario (Non-IID data, which is discussed later in the paper) while only slightly affecting the classification accuracy of the model. 
Therefore, we opt to use the standard \textit{FedAvg} instantiation.

\subsection{Attacks against FL}\label{preliminary:attacks}
Prior work shows that FL may be vulnerable to  attacks during and after the learning phase, targeting {\em robustness} %
and/or {\em privacy}%
~\cite{de2021critical,lyu2020threats}.

\descr{Poisoning Attacks.} These aim to make the target model misbehave; they can be performed either on the data or the model.
The former happens during the local data collection, while the latter occurs during model training. 
Poisoning attacks can be random or targeted; random ones reduce the utility of the aggregated FL model, while targeted attacks make the aggregated FL model output predefined labels. 

\descr{\em Backdoor Attacks.} A subclass of poisoning attacks, namely, backdoor attacks, has recently attracted a lot of attention from the research community~\cite{bagdasaryan2020backdoor,bhagoji2019analyzing}.
These are targeted model poisoning attacks where a malicious client injects a backdoor task into the final model, typically using a model-replacement methodology\cite{bagdasaryan2020backdoor,sun2019can}. 

As in~\cite{bagdasaryan2020backdoor}, %
the main steps of the distributed backdoor attack are as follows.
At round $r$, the attacker attempts to introduce a backdoor and replaces the aggregated model $\theta$ with a backdoored one $\theta^*$, by sending the following model update to the server:\vspace{-0.1cm}
 \begin{equation}\small
	\Delta\theta_{r}^{attacker} = \dfrac{\Sigma_{i=1}^{K}{n_i}}{\eta n_{attacker}} \cdot (\theta^* - \theta_r)%
   \end{equation}
where $n_i$ is \#data points at participant $i$, $\eta$ is the server learning rate, and the first term of the dot product is the boost factor.
Then, the aggregation in the next round yields:\vspace{-0.2cm}
\begin{equation}\small
	\Delta\theta_{r+1} = \theta^* + \eta\dfrac{\Sigma_{i=1}^{K-1}{n_i \Delta\theta_{r}^i}}{\Sigma_{i=1}^{K}n_i}\vspace{-0.1cm}
   \end{equation}
If we assume the training process is in its last rounds, then the aggregated model will converge; thus, model updates from non-attacker participants are small, and we would have $\Delta\theta_{r+1} \simeq \theta^*$. 
If the attacker does not know the learning rate or the number of training data hosted by the other participants, it can choose a large-enough boost factor, ensuring a good classification accuracy over backdoor samples.

Compared to untargeted data poisoning attacks, which aim to globally reduce the detection or prediction performances on any input, backdoor poisoning only triggers attack-desired misclassification over particular inputs embedded with a predefined trigger signal by the attacker. For other inputs, backdoor poisoning rarely introduces perturbation to the utility, thus yielding a stealthier threat to the integrity and the applicability of FL-based systems in practice~\cite{chen2017targeted,xieICML21,CaoJG21}. 
Attackers could use backdoor poisoning to mislead the FL system and produce the desired mispredictions over specific attack behaviors, e.g., causing security incidents of high priorities, while the system keeps working normally for attacks of low priorities. 
This could make the security event prediction model much less useful for organizations.

\descr{Inference Attacks.} These aim to exploit model updates exchanged between the participants and the central server to extract information about training data points.
The goal is to infer properties of these points that may be even uncorrelated with the main task or training set membership. 
In this paper, we focus on so-called {\em membership inference attacks} (MIA) and experiment with the two specific attacks discussed next.
 
\descr{\em Nasr et al.~\cite{nasr2019comprehensive}'s attack.} 
The main intuition is that each training data point influences the gradients of the loss function recognizably, i.e., a malicious participant can perform gradient ascent on a target data point before updating their local parameters.
If the point is part of a victim participant's set, the Stochastic Gradient Descent (SGD) algorithm reacts by abruptly reducing the gradient, and this can be recognized to infer membership successfully.
An adversarial participant can observe the aggregated model updates and, by injecting adversarial model updates, extract information about the union of the training dataset of all other participants.

\descr{\em Zhang et al.~\cite{zhang2020gan}'s attack.} %
This involves two main steps. (1) Augmenting the training data with Generate Adversarial Network (GAN), generating training data with the same distribution. %
The generator generates data records from random noise, and the discriminator is initialized with the target FL model. 
The target model as the discriminator can guide the generator to follow the training data points; the adversary then queries the target model with the generated samples and gets the labels. 
(2) Training a binary classification model using a GAN-enhanced attack method, aiming to differentiate members from non-members of other participants' training data.
Both the generated and the original data are used to train the attack model and predict training set membership.

\subsection{Defenses}\label{preliminary:robustness} 
We now briefly discuss state-of-the-art defenses against robustness attacks (namely, \textit{Trimmed Mean}, \textit{Krum}, \textit{FLTrust}, \textit{DnC}, \textit{Norm Bounding}, and \textit{Weak Differential Privacy}) as well as privacy attacks (namely, participant-level differential privacy). 
As shown in~\cite{DBLP:journals/corr/abs-2009-03561}, participant-level differential privacy can also protect robustness.

\descr{Trimmed Mean~\cite{yin2018byzantine}.}
For each model parameter, the server collects its values in all local model updates and sorts them. Given a trim parameter $\beta < n/2$, the server removes the largest and the smallest $\beta$ values and then computes the mean of the remaining $n-2\beta$ values as the value of the corresponding parameter in the global model update. The trim parameter $\beta$ should be at least the number of malicious clients to make Trim-mean robust. In other words, Trimmed mean can tolerate less than 50\% of malicious clients.

\descr{Krum~\cite{blanchard2017machine}.}
This defense assumes that the number of attackers is bounded and known; given the gradient updates from all clients at each iteration, malicious contributions will appear anomalous. 
The selection strategy by the server is to find one whose data is closest to that of other participants.
In other words, it computes the local sum of squared Euclidean distances to the other participants and chooses the one with minimal sum to update the global model.

\descr{FLTrust~\cite{cao2020fltrust}.}
The server collects a clean training set and maintains a model on such a dataset, denoted as the server model. 
At each iteration, it computes a trust score based on the deviation between local model updates and the server model. 
The server normalizes the magnitudes of the local model updates so that they lie in the same hyper-sphere as the server model update in the vector space. 
This limits the impact of malicious local model updates. %

\descr{DnC~\cite{shejwalkar2021manipulating}.} The idea behind poisoning attacks is that malicious updates are impactful in an IID FL setting if they shift from benign updates in a specific direction. 
The Divide-and-Conquer (DnC) defense computes the principal component of all updates, calculates the scalar product of the model updates with the principle component (called projections), and removes a fraction of the submitted model updates with the largest projections.

\descr{Norm Bounding/Weak DP~\cite{sun2019can}.}
Using boosted attacks for introducing the backdoor is likely to produce updates with large norms.
Therefore, if model updates received from attackers are over some threshold, the server could simply ignore those participants.
However, if the attacker is aware of the threshold, it can return updates within that threshold.
With norm bounding~\cite{sun2019can}, the idea is to guarantee that the norm of each model update is small even if the threshold is known.
In other words, if we assume that the updates' threshold is $T$, then the server can ensure that the norms of participants' updates are within the threshold by aggregating the model updates as follows:\vspace{-0.15cm}
 \begin{equation}
 \small
	\Delta\theta_{r+1} {=} \sum_{i=1}^{k}{\dfrac{\Delta\theta_{r+1}^{k}}{\max\left(1,\frac{{\| \Delta\theta_{r+1}^{k}\|_2 }}{T}\right)}}%
   \end{equation}
Weak Differential Privacy (WP)~\cite{sun2019can} can also be used as an additional defense; i.e., besides norm bounding, the server also adds Gaussian noise, further reducing the effect of poisonous data.

\descr{Central Differential Privacy (CDP).}
In CDP, also known as participant-level DP, the server perturbs the aggregation function. 
This guarantees that the function's output is indistinguishable, with probability bounded by an $\epsilon$, to whether or not a given participant is part of the training process.
This bounds the vulnerability to inference attacks and, overall, to information leakage from the (aggregated) model updates.
Ostensibly, participants need to trust the server to perform perturbation by adding noise correctly. 
In our experiments, we follow a similar implementation of CDP as~\cite{mcmahan2017learning,geyer2017differentially} as presented in Algorithm~\ref{alg:cdpalgo}.

\newlength{\textfloatsepsave} \setlength{\textfloatsepsave}
{\textfloatsep} \setlength{\textfloatsep}{0pt}
\begin{algorithm}[t]
\footnotesize
	\DontPrintSemicolon
  \SetKwFunction{FMain}{Main}
   \SetKwFunction{FPUPDATE}{Participant\_Update}
  \SetKwProg{Fn}{Function}{:}{\Return}
  \Fn{\FMain{}}{
		Initialize: model $\theta_0$, Moment\_Accountant($\epsilon$, N)\tcp*{N = \#participants}
        \For{each round $r=1,2,...$}{
				$C_r\gets $ randomly select participants with probability q\;
				$p_r\gets $Moment\_Accountant.get\_privacy\_spent() \tcp*{Returns privacy budget spent for current round}
				\If{$p_r > $ $T$\tcp*{If privacy budget spent greater than threshold, return current model}}{ 
					return $\theta_r$\;
				}
				\For{each participant $k \in C_r$} {
					$\Delta_k ^ {r+1} \gets$Participant\_Update({$k, \theta_r$})\tcp*{Done in parallel}
		
				}
	$S\gets {bound}$\;
	$z\gets {noise\_scale}$\;
	$\sigma\gets {zS/q}$\;
	$\theta_{r+1}\gets {\theta_r + \Sigma_{i=1}^{C_r}{\Delta_i ^ {r+1} }/C_r +  N (0, I\sigma^2)}$\;
	Moment\_Accountant.accumulate\_spent\_privacy($z$)\;

        }
        }
  \;
  
  \SetKwProg{Pn}{Function}{:}{\Return$\theta-\theta_r$\tcp*{This one is already clipped}}
  \Pn{\FPUPDATE{$k$, $\theta_r$}}{
  		$\theta\gets {\theta_r}$\;
  		\For{each local epoch $i$ from 1 to E}{
    				\For{batch $b \in B$}{
    					$\theta \gets \theta - \eta\nabla L(w;b)$\;
					$\Delta \gets \theta - \theta_r$	\;		
					$\theta \gets \theta_0 + \Delta \min\left(1,\dfrac{S}{\| \Delta\|_2 }\right)$\;
    				}
    		}
  }
	\caption{Central Differential Privacy in FL.}\label{alg:cdpalgo}
  	\vspace{-0.4cm}
\end{algorithm}

\setlength{\textfloatsep}{10pt}

\section{\systemName: Federated Prediction of Security Events}\label{sec:flachitecture}

\begin{figure}[t]
 	\centerline{
 	\includegraphics[width=1.0\linewidth]{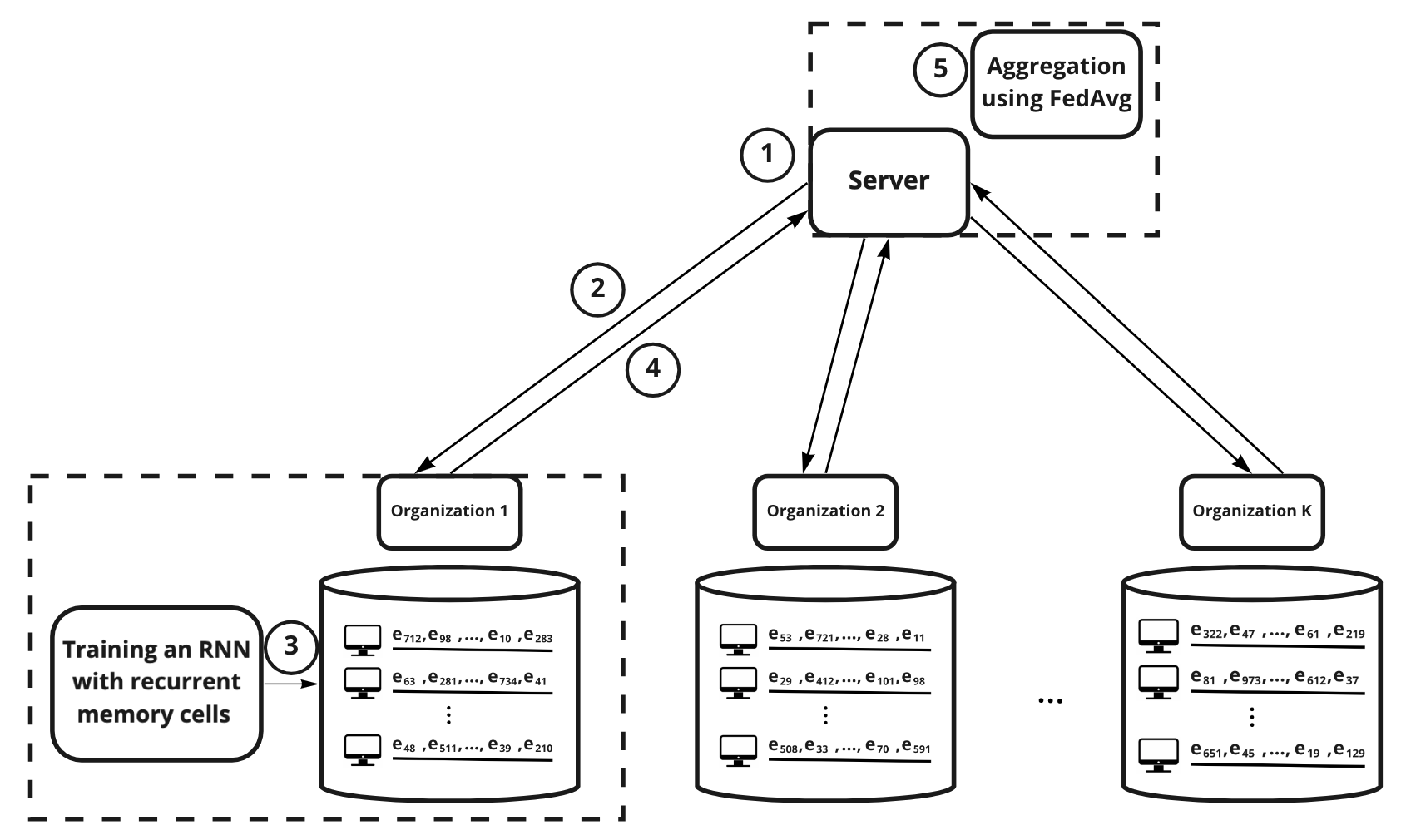}}
\caption{Workflow of \systemName.}%
	\label{fig:cerberusflow}
\end{figure}

We now present \systemName, a system supporting the federated prediction of security events. %
In a nutshell, the system involves a central server, which mainly takes care of parameter aggregation, as well as a number of organizations (or {\em participants}), each training a recurrent neural network (RNN) model with the same model architecture.
In the rest of this section, we describe \systemName's various entities and components.
Fig.~\ref{fig:cerberusflow} outlines the overall workflow of \systemName as per the following steps:

\begin{enumerate}
\item At each round, the server selects a fraction of the participating organizations for federated training of the prediction model. 
\item  The server sends the parameters of the RNN-based prediction model, aggregated at the server, to the organizations (parameters are initialized at random in round one).
\item Organizations update the RNN model with recurrent memory cells using stochastic gradient descent and the local training data based on the received model parameters.
\item The organizations send back the updated RNN model parameters to the central server.
\item The server aggregates the local model parameters sent by the selected organizations using FedAvg to produce the new global model parameters.
\end{enumerate}

\subsection{Components}
\label{sec:components}
\systemName consists of the following components:

\descr{Organization.} We operate in a collaborative setting with a number of organizations federating and engaging with \systemName to train a federated RNN model geared to predict security events. 
These correspond to participants in the traditional FL notation.

\descr{Security Event. }A security event $e_i$ is a timestamped observation at timestamp $i$. 
Examples of security events can be grouped into two categories: (i) system-level security events such as Adware.TopMoxie Activity, Trojan.Pandex Activity, etc., and (ii) network-level security events like TCP Bot Traffic Activity, HTTP IIS WebDav Remote Authentication ByPass, etc.

\descr{Machine.} Each organization may include several machines as computing devices.
These host an intrusion protection product, which generates a sequence of security events.
The sequence is ordered by timestamps. 

\descr{Server.} The central server is the entity in \systemName responsible for collecting local models, aggregating them, and sending the updated global model back to organizations.
The aggregation is performed using FedAvg as presented in~\ref{preliminary:fl}.
The participating organizations trust the server to exchange model parameters.

\descr{Model.} The ultimate goal of \systemName is to train a recurrent neural network (RNN) model that learns a sequence prediction function. 
More specifically, we follow the same approach as \textit{Tiresias}~\cite{shen2018tiresias}.
The model accepts a historical variable-length sequence of security events and predicts the future event:  	\vspace{-0.2cm}
\begin{align*}
  f : {e_{i}, e_{j}, ..., e_{k}} \to e_{target}\vspace{-0.1cm}
\end{align*}

\subsection{Training}
\label{sec:training}
\begin{figure}[t]
 	\centerline{
 	\includegraphics[width=0.98\linewidth]{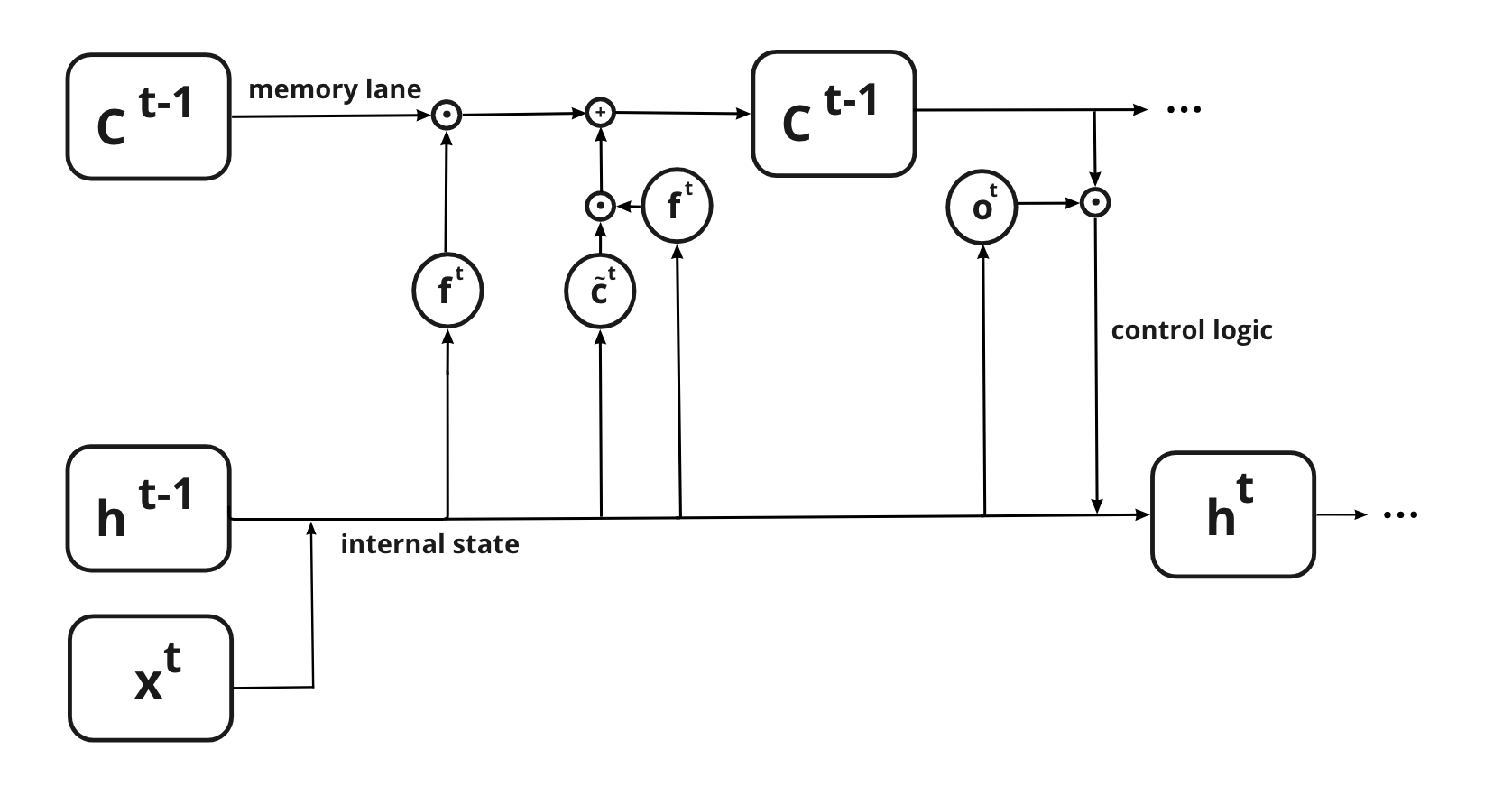}}
\caption{Single time-step update of the recurrent memory array.}
	\label{fig:memoryupdate}
\end{figure}

Training the model is conducted locally at the organization side, more specifically, by training an RNN with recurrent memory cells. 
Instead of stacked RNNs~\cite{shen2018tiresias}, which lack the generalization to new data, we use recurrent memory arrays~\cite{rocki2016recurrent} to build more complex memory structures inside an RNN cell.
According to~\cite{chen2020aistats}, the recurrent memory array-based model has a tighter generalization bound compared to the stacked RNN. 
Consistently, as shown by experimental observations in~\cite{RobertoArXIV2021}, the recurrent memory array outperforms LSTM (a stacked form of RNNs) in practice. 
We follow~\cite{rocki2016recurrent,shen2018tiresias} and define the recurrent memory array with the step update presented in Fig.~\ref{fig:memoryupdate} using the  following six equations:

\begin{enumerate}
\item $f^t = \sigma(W_fx^t + U_fh^{t-1} + b_f) $
\item $i^t = \sigma(W_ix^t + U_ih^{t-1} + b_i) $
\item $o^t = \sigma(W_ox^t + U_oh^{t-1} + b_o) $
\item $\tilde c^t = tanh(W_cx^t + U_ch^{t-1} + b_c) $
\item $c^t = f_t \odot c^{t-1} + i_t \odot \tilde c^t $
\item $h^t = o_t \odot tanh(c^t) $
\end{enumerate}
$W$ and $U$ are the matrices of input and hidden states.
$x$ and $o$ are the input and output.
$c$, $h$, and $f$ indicate cell state, hidden state, and forget state.
$\odot$ is an element-wise multiplication. 

This preserves temporal memories between successive security events for better generalization and maintains computational efficiency because of the single-layer RNN network.
Once the aggregated model converges, it takes a series of historical security events as input and predicts the probability distribution over the possible event types in the future. 
We use the log index (an integer number) of the security events as input to the RNN prediction model.
In this sense, we consider the type of security events as the categorical feature of each security event.

\systemName accepts variable-length security event series and predicts the target event.
The predicted output is the future security event with the highest prediction probability.

Unlike~\cite{shen2018tiresias}, \systemName does not have the performance monitoring phase. 
Our work is orthogonal to whether the monitoring module is involved, as  it is a plug-in component in~\cite{shen2018tiresias}.
Adopting variable or fixed-length inputs depends on the data format of different concrete prediction tasks. 
CERBERUS can take both formats as inputs; however, discussing the impact of the length is beyond  the scope of the present work.

\section{Building and Analyzing Distributed Datasets of Security Events}\label{sec:dataset}
\systemName is a {\em generic} framework for predicting security events.
To evaluate its performance in a real-world use case, we use security events collected by a major security company's intrusion prevention product. %
We call this the {\em original} dataset (see Sec.~\ref{sec:original}). 

We distribute this dataset to different participants in the FL system and analyze different distribution settings of this dataset to reason on the viability of using FL for the task at hand. %
We do so since, in typical FL settings, the training data of each participant might not always be Independent and Identically Distributed (IID).
In fact, the heterogeneously distributed local training data sets may severely affect the performance of the federated model aggregation; this is known as the \textbf{\em Non-IID issue} in FL~\cite{mcmahan2017communication}. 
In practical applications, security events from different organizations %
can vary significantly due to different security postures, hardware settings, and different security policies they may enforce.
As a result, it is crucial to evaluate the effectiveness of distributed training over Non-IID learning tasks.

To quantify how non-independent and non-identical the dataset distribution is, we define and use the Non-IIDness score (see Sec.~\ref{sec:non-iidness}).
We then consider synthetic scenarios to simulate the case for distributions with different levels of Non-IIDness. 
Overall, our goal is to evaluate the impact of increasingly more skewed (more "non-iid") local training datasets over the utility of the jointly trained security event prediction model (see Sec.~\ref{sec:distributions}). 

\subsection{{\em Original} Dataset}\label{sec:original}
Our starting point is a dataset of security events collected from a major security company's intrusion prevention product for 7 consecutive days in July 2021. 
We denote this dataset as the {\em original} dataset. 
The collected security events include network-level or system-level activities matching predefined firewall signatures from the security company (e.g., a network activity matching the heartbleed CVE-2014-0160 signature). 
For each security event, we collect the following metadata: the machine ID of the device reporting the event, the timestamp of the event, the security event ID designated by the security company, a short description of the event, and the system actions. 
Note that the events are recorded, and contributed by, users who explicitly opt-in to share data to help the security service provider improve the capabilities of detecting malicious incidents. 
\systemName learns a sequence prediction function that accepts a sequence (with either variable or fixed length) of the security events from the past as the input and predicts the upcoming event in the future.
In contrast to the previous approaches~\cite{Sabottke2015usenix,liu2015cloudy}, \systemName does not need to extract any additional features from the security events.

The dataset is partitioned into $\numberoforgs$ organizations, and we consider each organization as one participant in the FL system.
There are $\totalnumberofmachines$ machines in total, with each machine hosting a series of security events that are identified by a unique ID. 
In this paper, we identify each prediction event ID as the target event, i.e., the last event in the series, as a class.
There are $\numberofeventstypes$ classes which is also the number of different event types, and a total of $\totalnumberofevents$ events.

\subsection{Non-IIDness Score}\label{sec:non-iidness}

\begin{algorithm}[t]

 \footnotesize
	\DontPrintSemicolon
  \SetKwFunction{FMain}{Compute\_Non-IIDness\_Score}
   
  \SetKwProg{Fn}{Function}{:}{\Return$score$}
  \Fn{\FMain{}}{
				$N\gets $ all participants\;
				$histogram\_list \gets$ empty(); %
				
				\For{each participant $k \in N$} {
					\For{each class $c \in k.getDataset()$} {
						$temp = count\_number\_of\_training\_instances(c)$\;
						$hist = create\_histogram(temp)$\;
						$normalize(hist)$\; 
						$histogram\_list.add(hist)$\;
					}
				}
				
			$score = KL\_divergence(histogram\_list)$
        }
  \;
  	\vspace{-0.2cm}
	\caption{The Non-IIDness score of an FL distribution.}\label{alg:iddness}
\end{algorithm}

To quantify the Non-IIDness level of an FL distribution, we use Algorithm~\ref{alg:iddness}, which returns a numerical value, denoted as the Non-IIDness score.
The algorithm works as follows:
(1) First, we build a histogram per participant showing the occurrence frequency of different classes in the training data.
Each bin corresponds to one class. 
(2) Then, we normalize the histograms. 
(3) We compute the Kullback Leibler (KL) divergence score~\cite{Joyce2011} between the histograms. 
We take the average of the KL scores between every pair of organizations and set it as the Non-IIDness score.

Note that the KL divergence score quantifies how much the class distribution differs from one participant to another (if KL equals 0, the class distributions of two participants are equal).
Thus, the lower the average KL average score between every pair of participants is, the more IID the distributions among different participating organizations are.

\subsection{Distributions}\label{sec:distributions}
Besides the original dataset collected from the intrusion prevention product which forms the {\em{primary}} distribution, we simulate additional scenarios accounting for distributions with different Non-IIDness levels (called {\em{knowledgeable participant}} and  {\em{extreme Non-IID}}).
The knowledgeable participants distribution sheds light on organizations that contribute more to the system and how they might benefit.
The extreme Non-IID setting lets us study the lowest utility we can get.
To do so, we pool the data and distribute them with different Non-IIDness settings across multiple participants.
In the following, we present the setting of each distribution. 

\begin{itemize}
\item[$\bullet$] \textbf{Primary Distribution}: We consider all the organizations of the main dataset as participants of the FL setting resulting in 5,419 participants. 

\item[$\bullet$] \textbf{Knowledgeable Participants Distribution}: We consider participants having instances from {\em all} of the classes.
These are likely to contribute to the aggregate model's utility, as opposed to participants that rarely contribute to the training process.
We set the total number of participants to 2,000.
Each security event sequence has a target event. There are 1,456 different event types. If all security sequences are partitioned based on the target security event, we get 1,456 partitions.
We randomly sample a sequence without replacement from each partition and allocate it to each knowledgeable participant. We continue this process until all the security event sequences are allocated to knowledgeable participants.

We denote the percentage of knowledgeable participants out of the 2000 participants with $m$ , initially setting $m=60\%$.
Later in our experiments, we work with several values of $m$ to vary the impact of increasingly more knowledgeable participants.
The (1-m)\% of the participants are randomly selected from the participants of the main distribution with the number of classes less than (1,456/2)=728.

\item[$\bullet$] \textbf{Extreme Non-IID Distribution}: One class is assigned to each participant. 
Recall that a class is one unique type of security event predicted by a machine. 
As there are $\numberofeventstypes$ classes in total, we end up with $\numberofeventstypes$ participants, and security events related to their class are assigned to that participant. 
This way, each participant has a different data distribution, thus yielding an extreme Non-IID distribution.
 
\end{itemize}

\begin{table}[t]
\small
\begin{center}
\begin{tabular}{l@{}rr}
\toprule
 & \textbf{Non-IIDness}\\ 
             \textbf{Distribution}  & {\bf Score} & \textbf{\#Participants}
              \\  \midrule
Primary    & 3.42    &5,419     \\ 
Knowledgeable Participant & 9.61 	&2,000	  \\ 
Extreme Non-IID & 18.23   &1,465     \\ 
\bottomrule
\end{tabular}
\end{center}
\caption{Distributions' IID-ness Scores.} %
\label{tab:iidnessmeasurement}
\vspace{-0.2cm}
\end{table}

\begin{table}
\small
\begin{center}
\setlength{\tabcolsep}{2pt}
\begin{tabular}{l@{}rrrrr}
\toprule
             \textbf{ }& \textbf{Precision}& \textbf{Recall}& \textbf{F1} & \textbf{~Accuracy}& \textbf{FPR} \\  \midrule
Non-FL (Centralized) & 0.84  & 0.82  & 0.83 & 0.85 & 0.19 \\  \specialrule{0.05pt}{1pt}{1pt}
Primary  & 0.69   & 0.71 & 0.70 & 0.78 & 0.27  \\ 
Knowledgeable Participant & 0.62 & 0.65 & 0.63 & 0.72 & 0.29 \\ 
Extreme Non-IID  & 0.53 & 0.57 & 0.55 & 0.65 & 0.32 \\ 
\bottomrule
\end{tabular}
\end{center}
\caption{Utility Measurement of \systemName}
\label{tab:utilitymeasurement}
\end{table}

\descr{Non-IIDness Scores.} In Table~\ref{tab:iidnessmeasurement}, we report the Non-IIDness scores of each setting.
Unsurprisingly, the extreme Non-IID distribution has the highest score (18.23), %
while the primary distribution has the lowest score (3.42) as the original dataset is relatively independent and identically distributed. 
The knowledgeable participant distribution has both kinds of participants, hosting data from all the classes and data from just a few classes; therefore, it has a value of 9.61, somewhere in between the other two.

\section{Utility}\label{sec:utility}
In this section, we experiment with different distributions and metrics to address the evaluation of the aggregated model performance in \systemName, 
as well as how much the participants contribute to and benefit from the system.  

\subsection{Experimental Setup}
The first step of our evaluation is to assess the viability of a collaborative learning approach to security event prediction based on FL and RNN.
In particular, we experiment with data from an intrusion prevention product obtained from a major security company\footnote{Complete details are omitted due to non-disclosure agreement.} and use the \systemName system (see Section~\ref{sec:flachitecture}) trained over three distributions (see Section~\ref{sec:distributions}), with a varying number of participants.
We set the number of FL rounds to 200, while the number of local training epochs is 5. 
The participation rate (i.e., the number of participants selected on each round of the FL process) is set to 1 for all three distribution settings.
We do so to encompass a realistic use case of security incident prediction where it is likely that all participants can stay online during the FL training process and return local model updates consistently at each iteration.

Finally, out of $\totalnumberofmachines$ security series, we use 80\% of the data for training, 10\% for validation, and 10\% for testing. 

\subsection{Model Performance}
\label{modelperformance}
\descr{Centralized Approach Baseline.} We set a baseline for the prediction performance based on a centralized, non-federated version of the framework.
In essence, this provides us with an upper bound of the model performance.
The resulting precision, recall, F1 score, accuracy, and False Positive Rate (FPR) metrics are presented in Table~\ref{tab:utilitymeasurement} (top row).

Note that the performance is more or less comparable to Tiresias~\cite{shen2018tiresias}, on which our RNN-based model is based, and the difference is likely due to the different datasets at hand.\footnote{The implementation of Tiresias~\cite{shen2018tiresias} is not publicly available, so we re-implemented it from scratch.}

\descr{\systemName Performance.} We run \systemName on the three distributions and report performance metrics in Table~\ref{tab:utilitymeasurement}.
We implement a macro-average approach to compute the precision, recall, accuracy, and FPR independently for each class and then take an average (we treat all the classes of security events equally this way).

As expected, all the metrics, except FPR, are reduced compared to the baseline.
We provide a detailed discussion about FPR in Section~\ref{sec:discussionconclusion}.
The Non-IIDness of the distributions has an important effect on the aggregated model performance;
in fact, the lower the Non-IIDness score, the better the model is in terms of utility.

\begin{algorithm}[t]
 \footnotesize
  \SetKwFunction{FMain}{Compute\_Contribution\_Impact}
   
  \SetKwProg{Fn}{Function}{:}{\Return$M_k$;}
  \Fn{\FMain{Participant $k$}}{
				$M\gets $ aggregated model\;
				Remove $k$\;
				Retrain $M_k$\tcp*{$M_k$ is the aggregated model without $k$}
				Compute $M_k$'s precision\;				
				
        }
  \
	\caption{Computing\hspace{-0.032cm} a\hspace{-0.032cm} Participant's\hspace{-0.032cm} Contribution\hspace{-0.032cm} Impact.\hspace*{-1cm}}\label{alg:contributionscore}
\end{algorithm}

\begin{figure*}
\centering
\centering
\subfloat[Primary]{\label{influencescoreprimary}\includegraphics[width=0.33\linewidth]{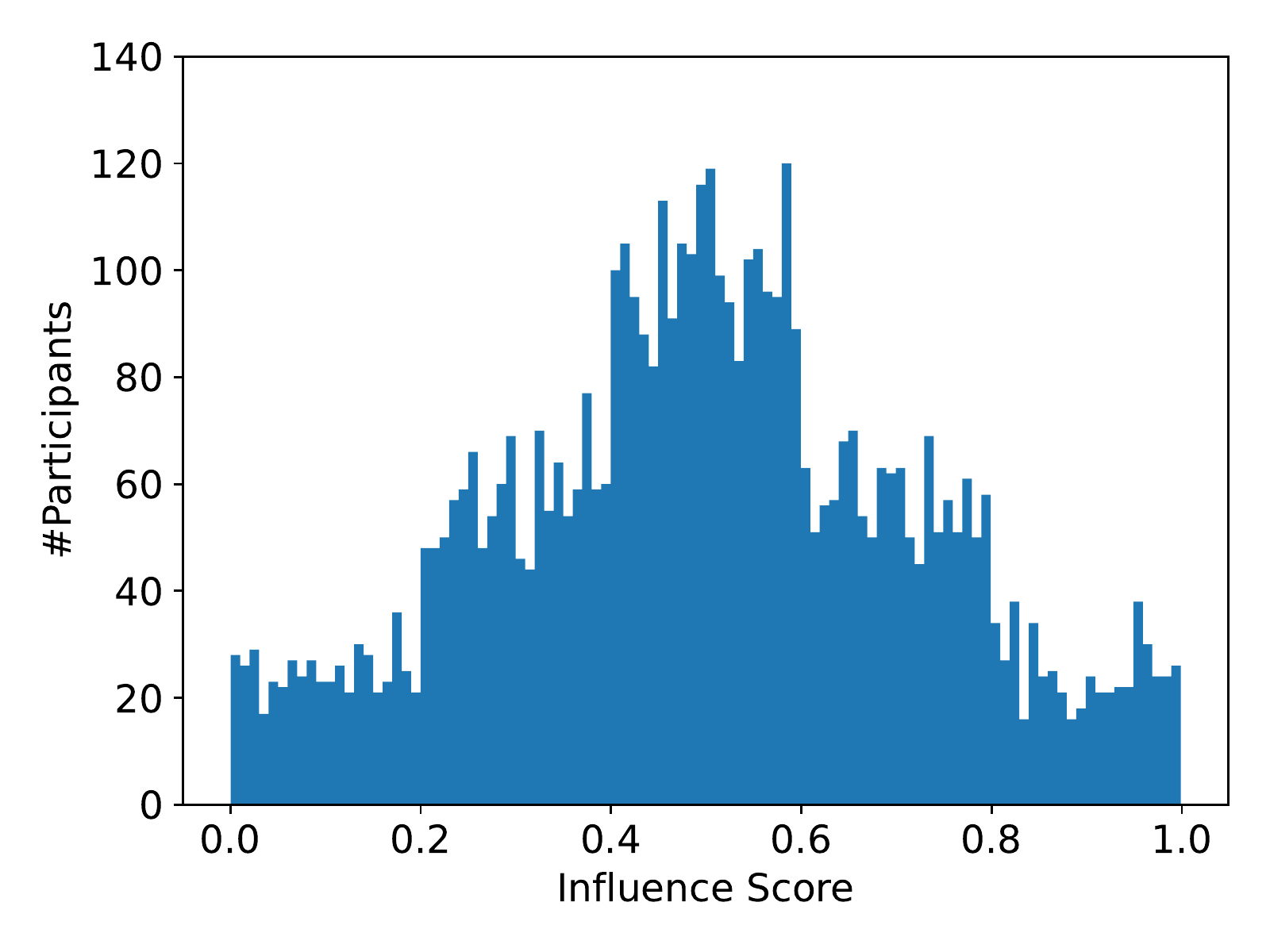}}\hspace*{-0.1cm}
\subfloat[Knowledgeable Participant]{\label{influencescoreknowledgeable}\includegraphics[width=0.33\linewidth]{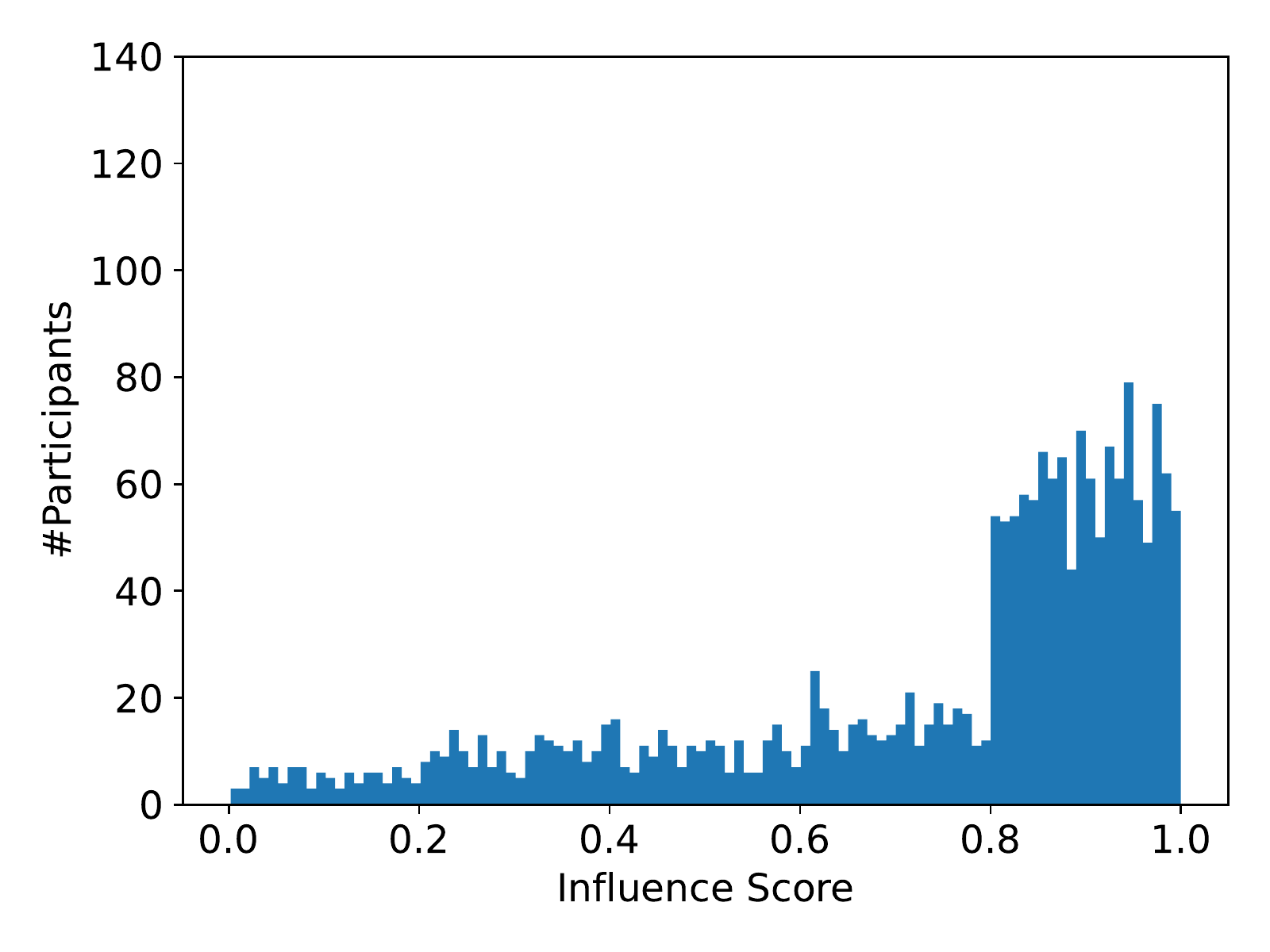}}\hspace*{-0.1cm}
\subfloat[Extreme Non-IID]{\label{influencescoreextreme}\includegraphics[width=0.33\linewidth]{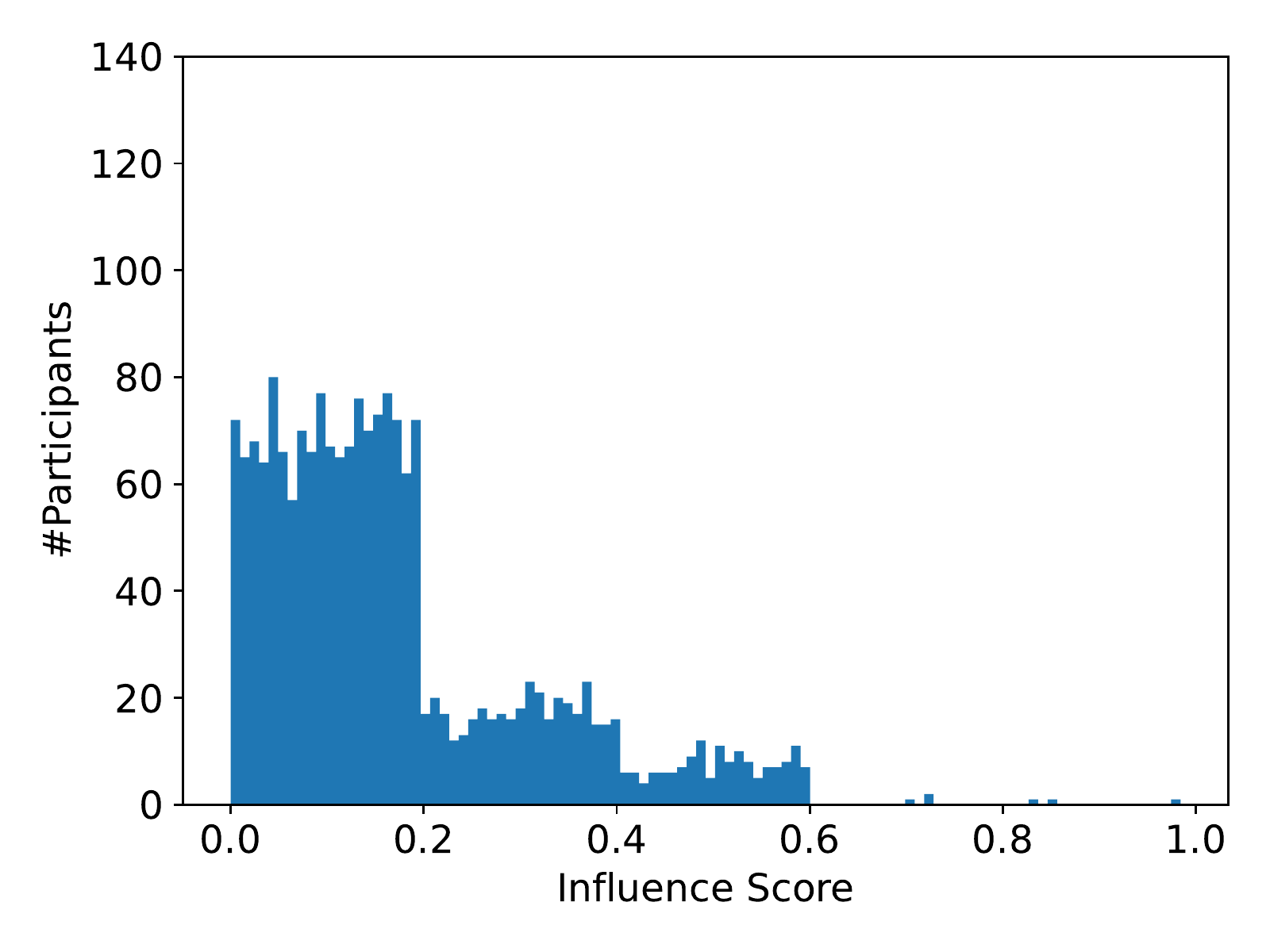}}\hspace*{-0.1cm}
\caption{Influence score distribution of Primary, Knowledgeable Participant, and Extreme Non-IID distributions.}
\vspace{-0.6cm}
\label{fig:influencescore}
\end{figure*}

\begin{figure*}
\centering
\centering
\subfloat[Primary]{\label{influencetop20primary}\includegraphics[width=0.33\linewidth]{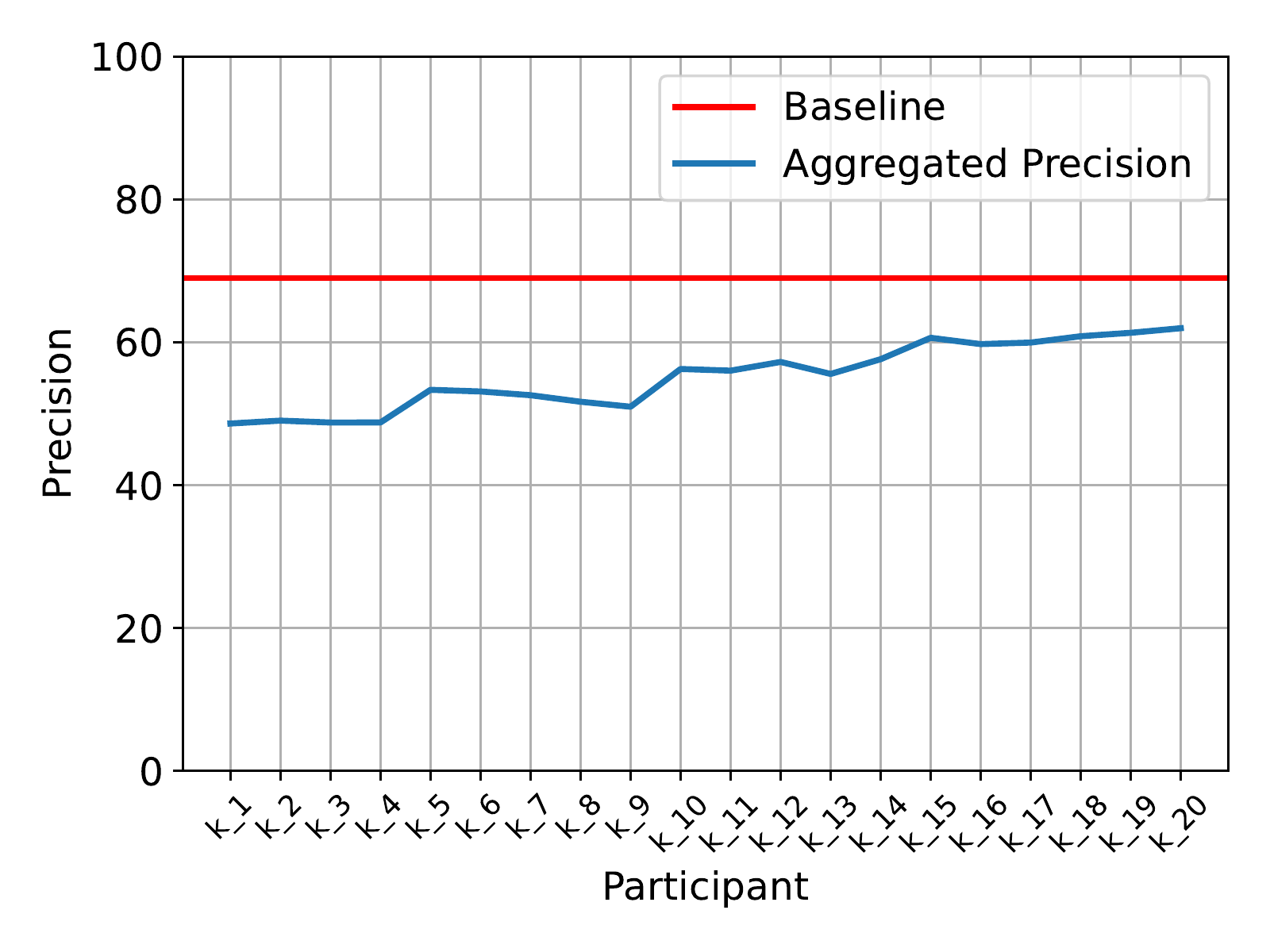}}\hspace*{-0.1cm}
\subfloat[Knowledgeable Participant]{\label{influencetop20knowledgeable}\includegraphics[width=0.33\linewidth]{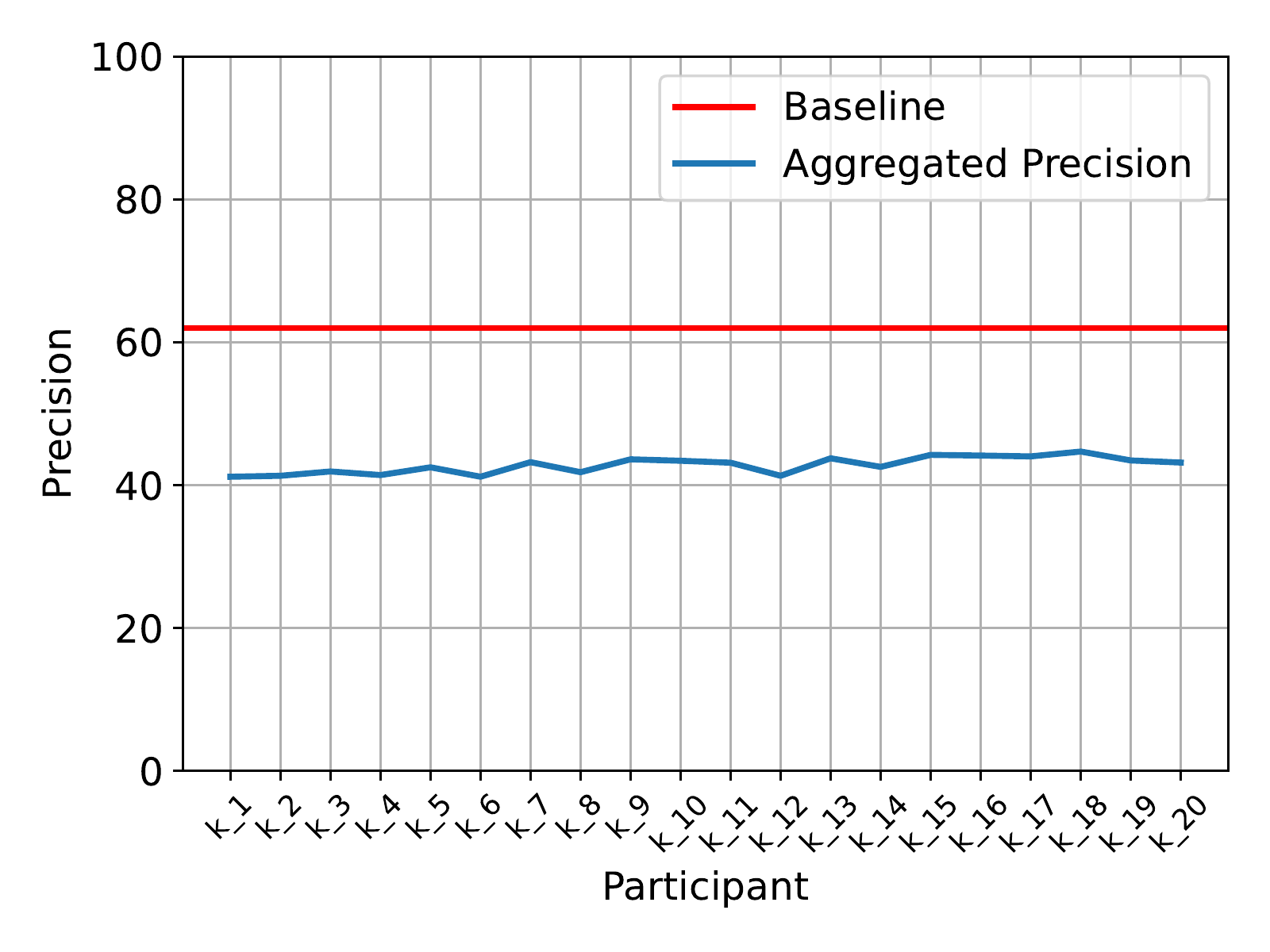}}\hspace*{-0.1cm}
\subfloat[Extreme Non-IID]{\label{influencetop20extreme}\includegraphics[width=0.33\linewidth]{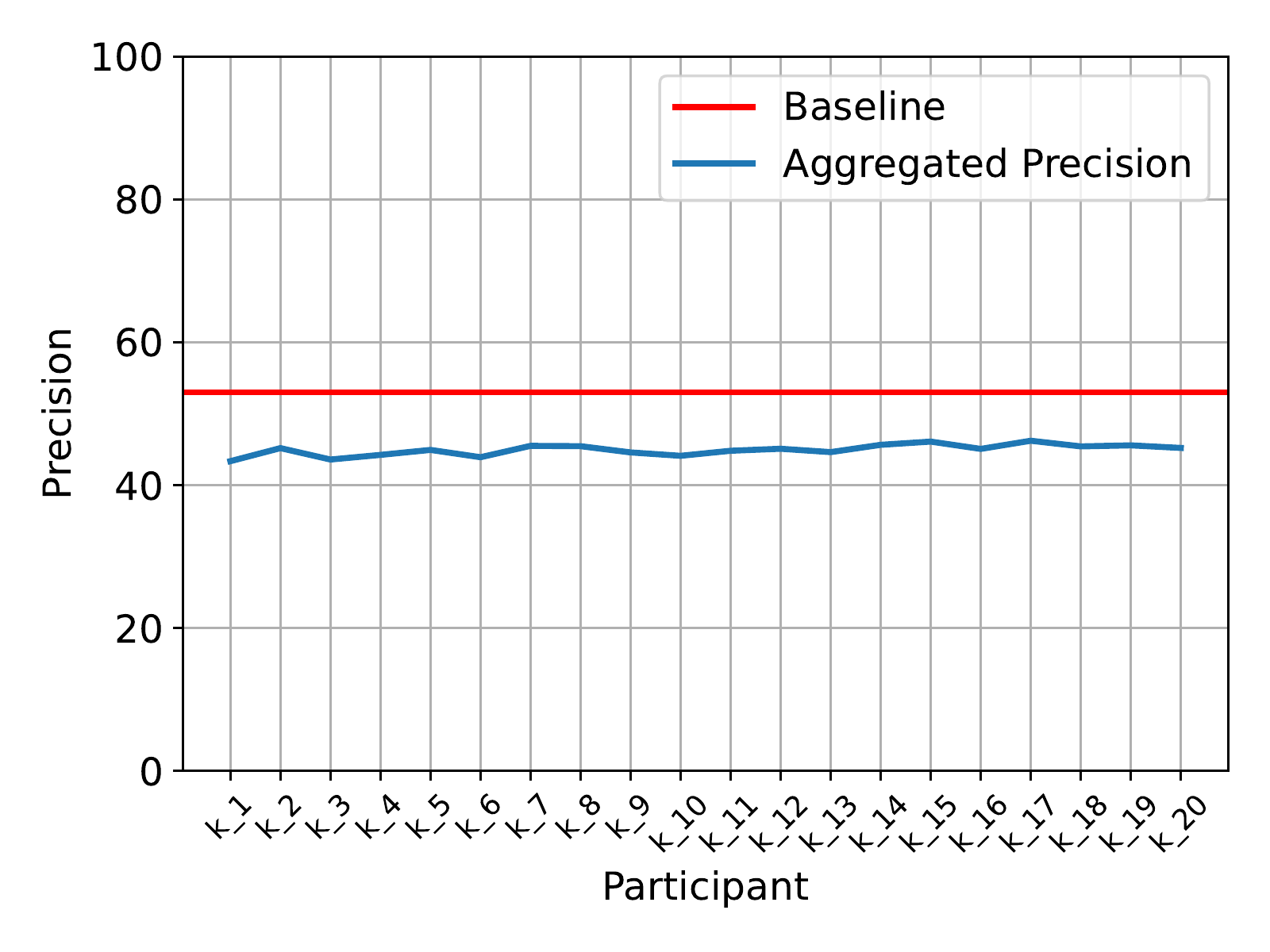}}\hspace*{-0.1cm}
\caption{Measuring the impact of top 20 impactful participants using Algorithm~\ref{alg:contributionscore}.} %
\label{fig:top20influencescorealgo}
\end{figure*}

\subsection{Participant's Contribution}

Next, we measure how participants in \systemName contribute to the system and the aggregated model.
We model FL as a cooperative game with updates from organizations as players and the model utility on the server's test dataset as the characteristic function. 
That gives us a Shapley value-based user importance scoring system~\cite{LundbergNIPS2017}.
We call this metric the \textbf{\em Contribution Impact} and compute it as per Algorithm~\ref{alg:contributionscore}.
In a nutshell, we remove each organization and measure the utility of the aggregated model. 

However, doing so for all participants would be computationally expensive; prohibitively so.
Arguably, the next best thing is to compute it for ``important'' participants.
To this end, we follow the same approach as~\cite{koh2017understanding} by using \textbf{\em Influence Functions}. 

Overall, the goal is to understand the effect of training points on a model's predictions, specifically, formalizing it as two questions: 1) how would the model's predictions change if we did not have this training point? 
2) how would the model's predictions change if a training input were modified?
Influence functions are asymptotic approximations of leave-one-out retraining under the assumption that the model parameters minimize the empirical risk and that the empirical risk is twice-differentiable and strictly convex.
As done in~\cite{agarwal2016second,koh2017understanding}, we use stochastic estimation to avoid iterating all training points, sampling a single point per iteration, and speeding up the process.
The influence score ranges between 0 and 1; the higher the value, the more influential the participant's training dataset.

We measure the influence score of the participants for the three settings of the data distributions, and the histogram plots are presented in Fig.~\ref{fig:influencescore}.
We observe that Fig.~\ref{influencescoreprimary} and Fig.~\ref{influencescoreknowledgeable} follow a normal distribution.
The Knowledgeable Participant distribution tends to have higher influence scores on average.
This is due to the presence of participants with training data instances of all classes. 
In the Extreme Non-IID distribution, most participants have low influence scores as each has data instances of only one class.
To evaluate the impact of a participant on the aggregated model and to verify the consistency between the contribution impact and influence score, we compute the contribution impact of the participants with the highest influence score. 
We run Algorithm~\ref{alg:contributionscore} on the top 20 organizations with the highest influence score, which we denote as \textbf{\em Impactful Participants}.

The results are reported in Fig.~\ref{fig:top20influencescorealgo}.
In the plots, the x-axis has the impactful participants sorted by their influence score.
The baseline represents the precision of the aggregated model server-side. 
The aggregated precision is measured when the specific impactful participant is removed from \systemName.
The gap between the two curves determines the contribution impact. 

In all three distributions, removing the impactful participants yields a significant drop in the aggregated precision. 
In Fig.~\ref{influencetop20primary}, the aggregation precision exhibits a descending trend as the influence score decreases, and hence the gap between the baseline and aggregated precision decreases. 
However, Fig.~\ref{influencetop20knowledgeable} and Fig.~\ref{influencetop20extreme} show that the aggregation precision trend stays the same. 
We believe this is due to:  1) All knowledgeable participants host training instances from all of the classes of security events; these knowledgeable and impactful participants tend to have a similar influence over the trained model. 
2) For the extreme Non-IID case, overlapping between different impactful participants is marginal due to the high Non-IIDness of local data distribution; thus, the impactful Non-IID participants have a similar influence over the trained model.

\begin{figure*}
\centering
\centering
\subfloat[Primary]{\label{top20localvsglobalprimary}\includegraphics[width=0.33\linewidth]{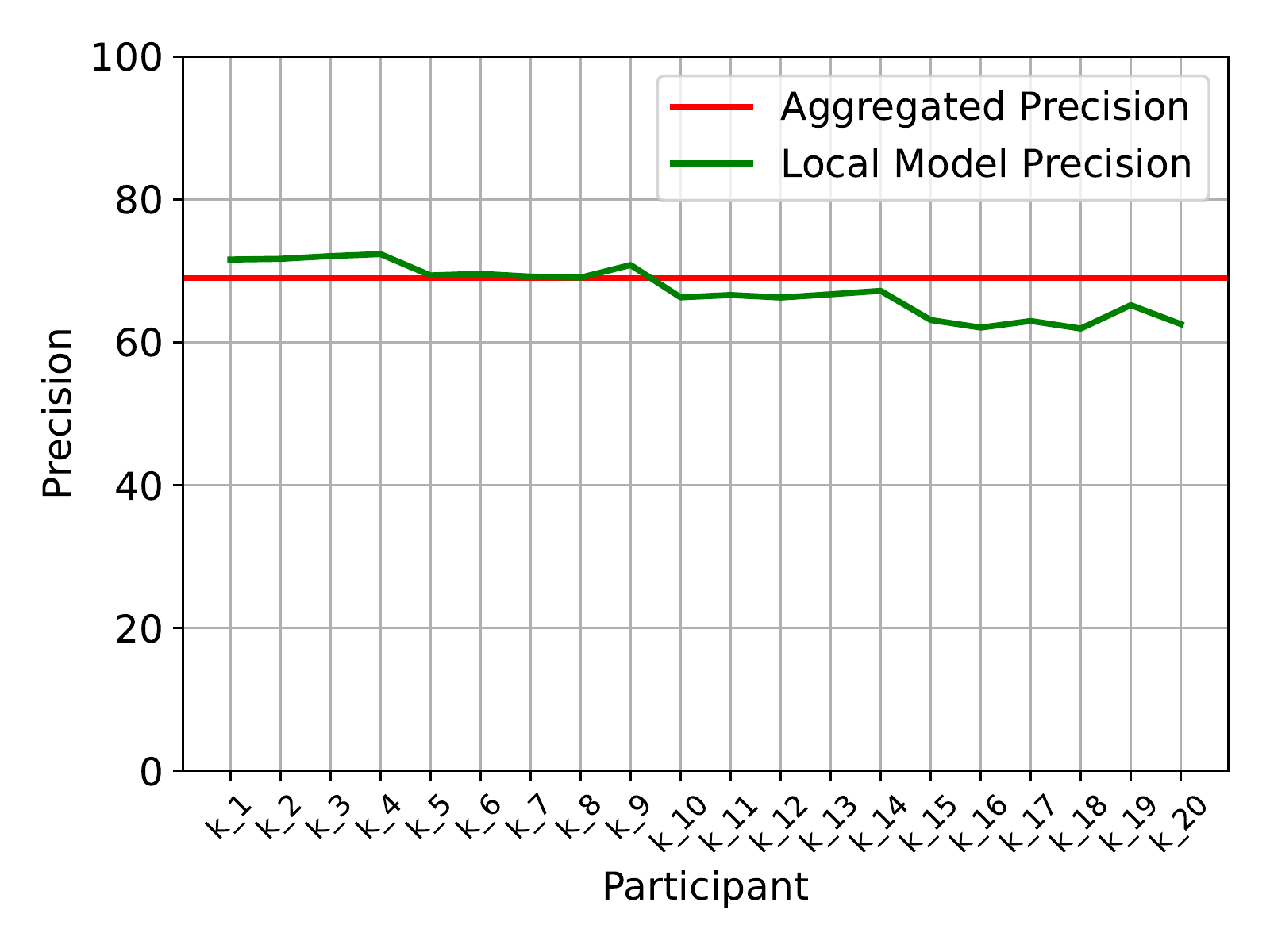}}\hspace*{-0.1cm}
\subfloat[Knowledgeable Participant]{\label{top20localvsglobalknowledgeable}\includegraphics[width=0.33\linewidth]{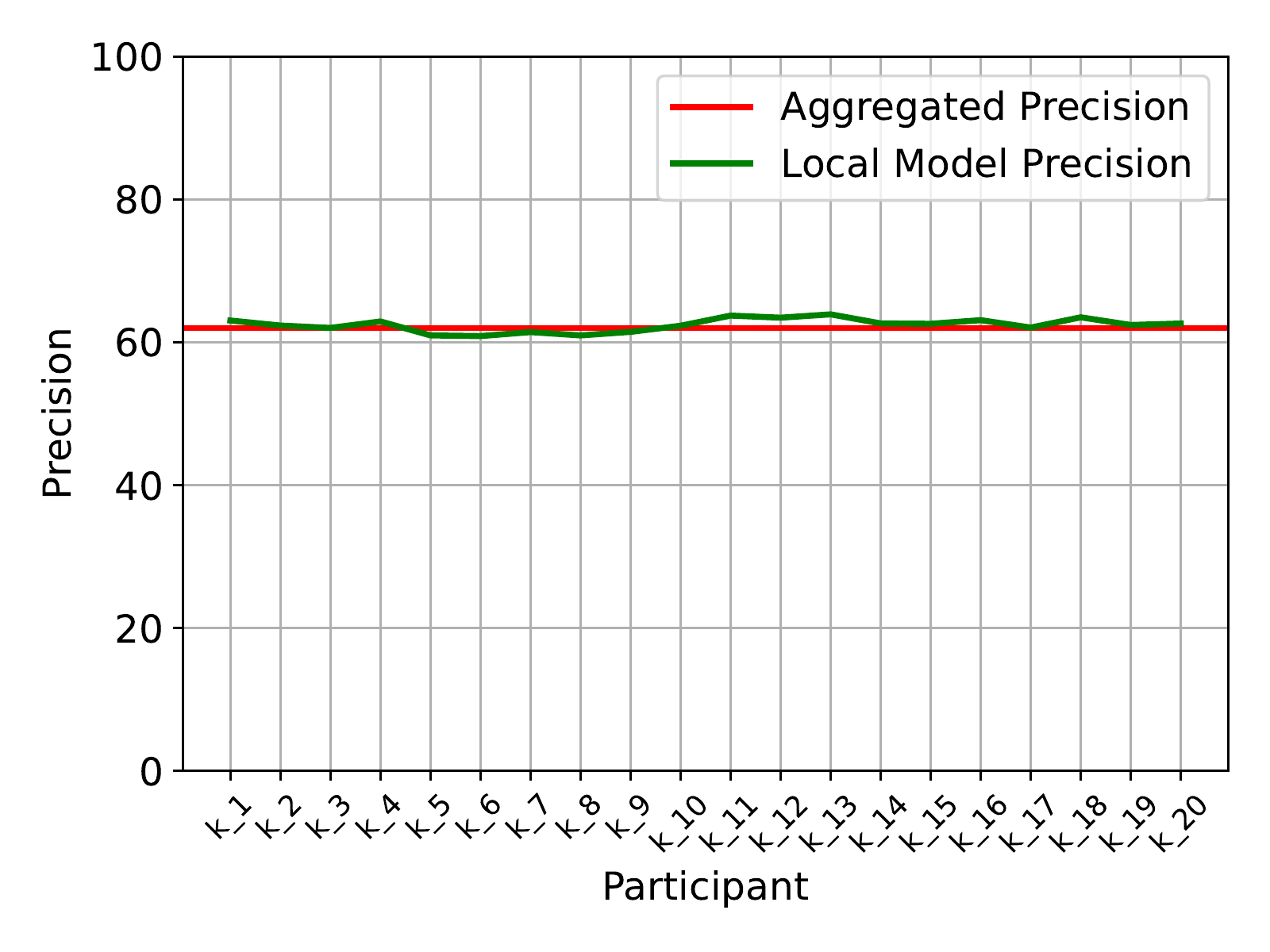}}\hspace*{-0.1cm}
\subfloat[Extreme Non-IID]{\label{top20localvsglobalextreme}\includegraphics[width=0.33\linewidth]{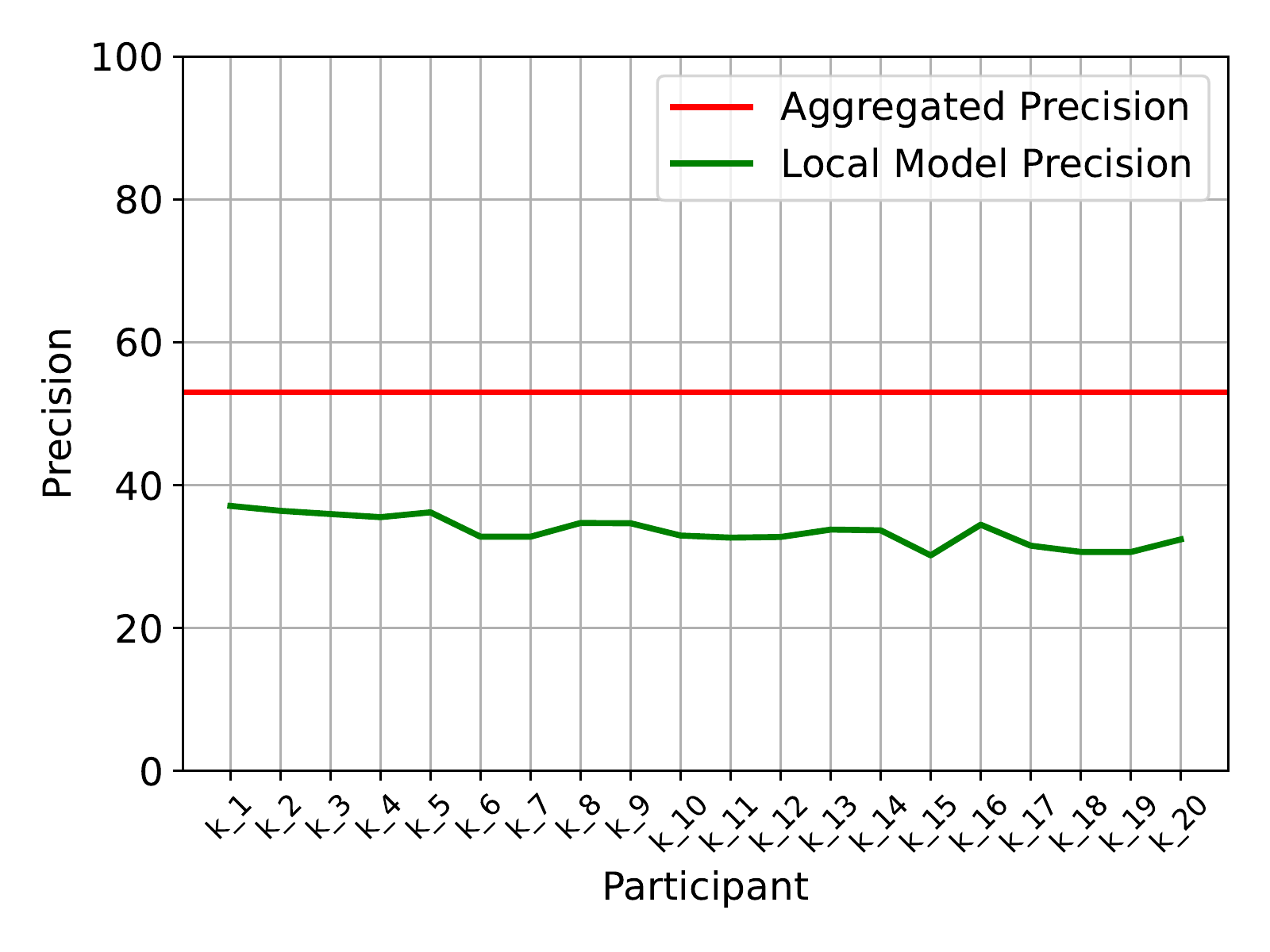}}\hspace*{-0.1cm}
\caption{Local vs aggregated model precision comparison for top 20 impactful organizations.}
\vspace{-0.5cm}
\label{fig:top20localvsglobal}
\end{figure*}

\begin{figure*}
\centering
\centering
\subfloat[Primary]{\label{totallocalvsaggregatedprimary}\includegraphics[width=0.33\linewidth]{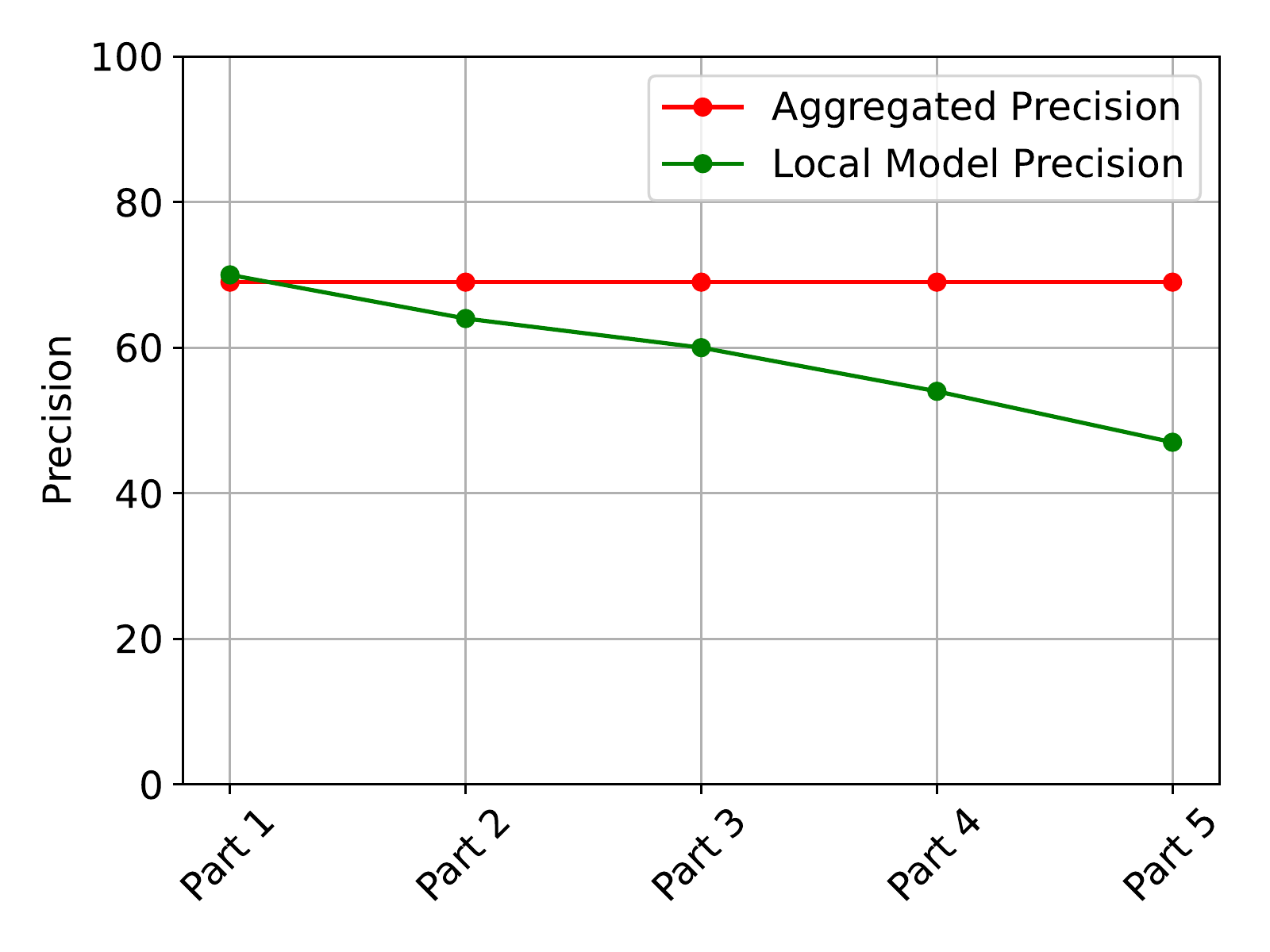}}\hspace*{-0.1cm}
\subfloat[Knowledgeable Participant]{\label{totallocalvsaggregatedknowledgeable}\includegraphics[width=0.33\linewidth]{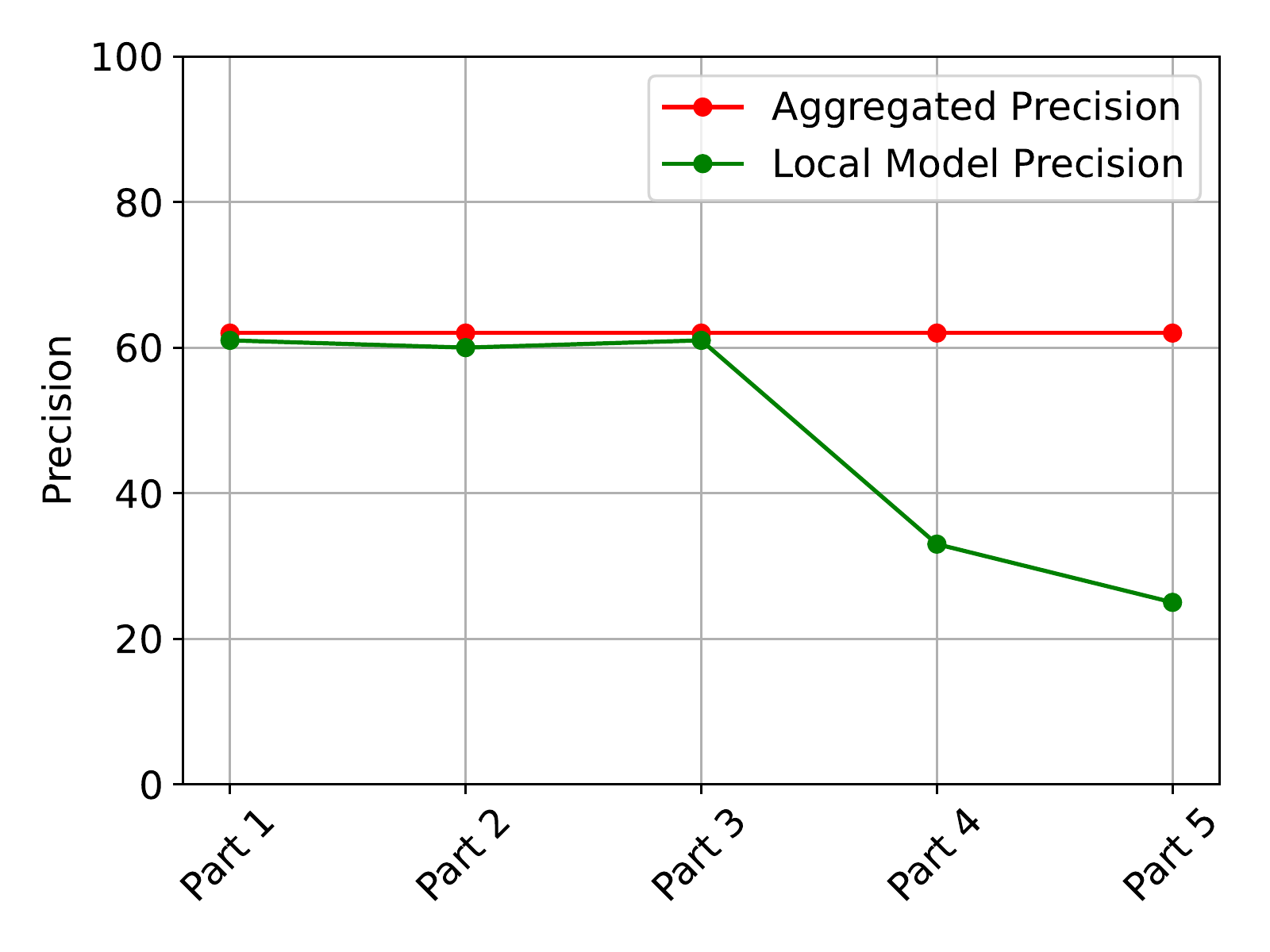}}\hspace*{-0.1cm}
\subfloat[Extreme Non-IID]{\label{totallocalvsaggregatedextreme}\includegraphics[width=0.33\linewidth]{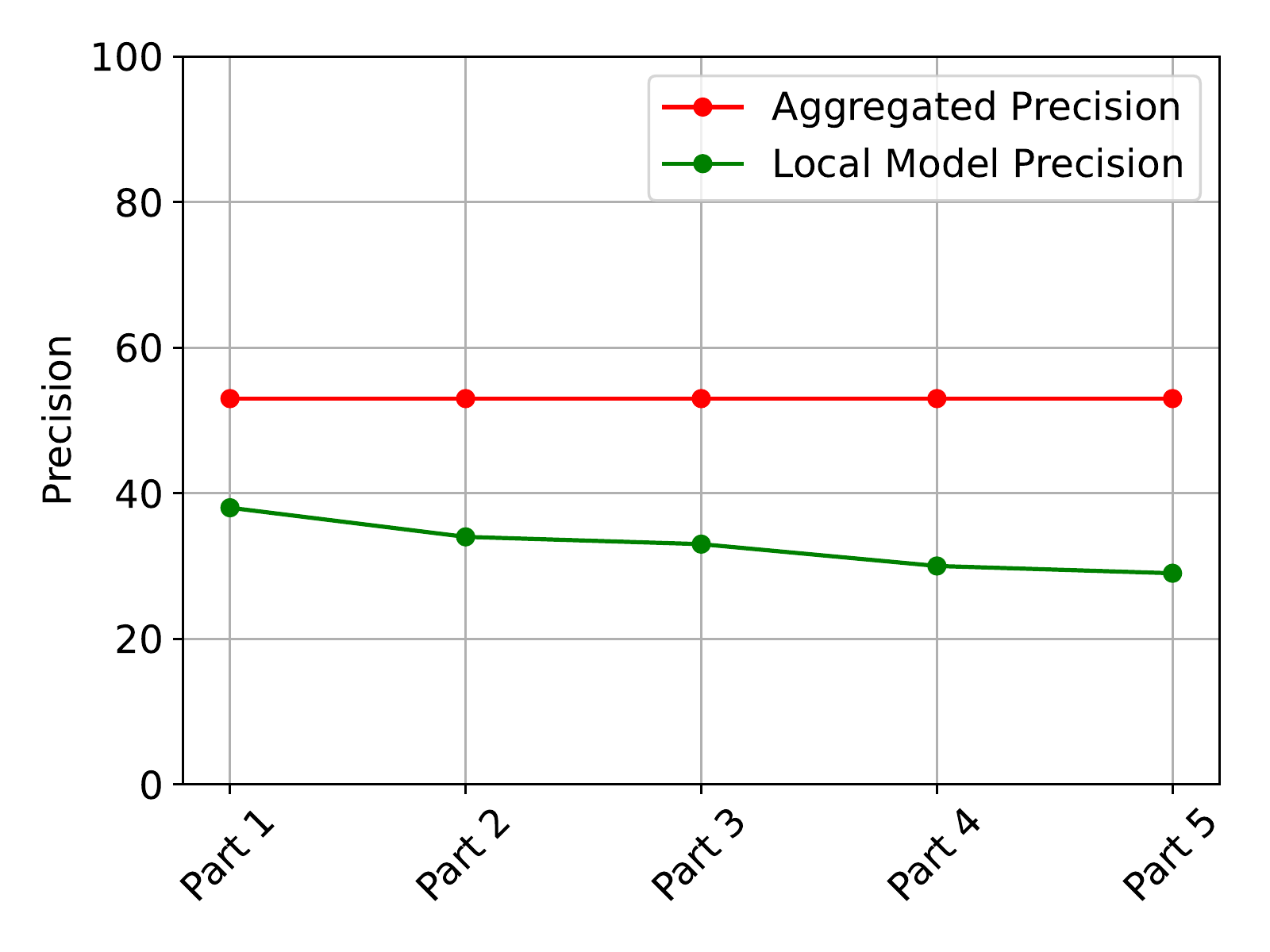}}\hspace*{-0.1cm}
\caption{Local vs aggregated model precision comparison for all organizations separated by 20 percent (every part consists of 20\% of total participants).}
\label{fig:totallocalvsaggregated}
\end{figure*}

\subsection{Participant's Benefit}
We then set out to investigate how organizations benefit from \systemName. 
To do so, we create a fixed held-out testing set that includes all the classes (i.e., security event types), which we denote as \textbf{\em Examination Test}. 
It includes 30,000 samples.
We do random sampling and choose 70\% of the data for training and the rest for testing.
Once the training phase in \systemName terminates, we evaluate each organization's local and aggregated models against this held-out testing set.

In other words, we scrutinize each organization as well as the aggregated model.
For instance, some organizations could report biased/noisy data resulting in biased local models;  this might also affect the aggregated model. 
To remove such potential bias, we repeat the experiment five times. 
We resample the examination test  at each round and retrain the aggregated model in \systemName. %

Fig.~\ref{fig:top20localvsglobal} reports the results for the top 20 impactful organizations.
From Fig.~\ref{top20localvsglobalprimary}, we observe a few participants with higher local precision than the aggregated one.
Fig.~\ref{top20localvsglobalknowledgeable} shows that most of the impactful participants have the same local model precision as the aggregated one.
However, Fig.~\ref{top20localvsglobalextreme} indicates that all the impactful participants have worse local precision than the aggregated model.
Furthermore, we perform the previous experiment for all the organizations, and the results are presented in Fig.\ref{fig:totallocalvsaggregated}. 
On the x-axis, each part consists of 20\% of the total participants and is sorted by decreasing influence score.

\begin{figure}
 	\centerline{
 	\includegraphics[width=0.7\linewidth]{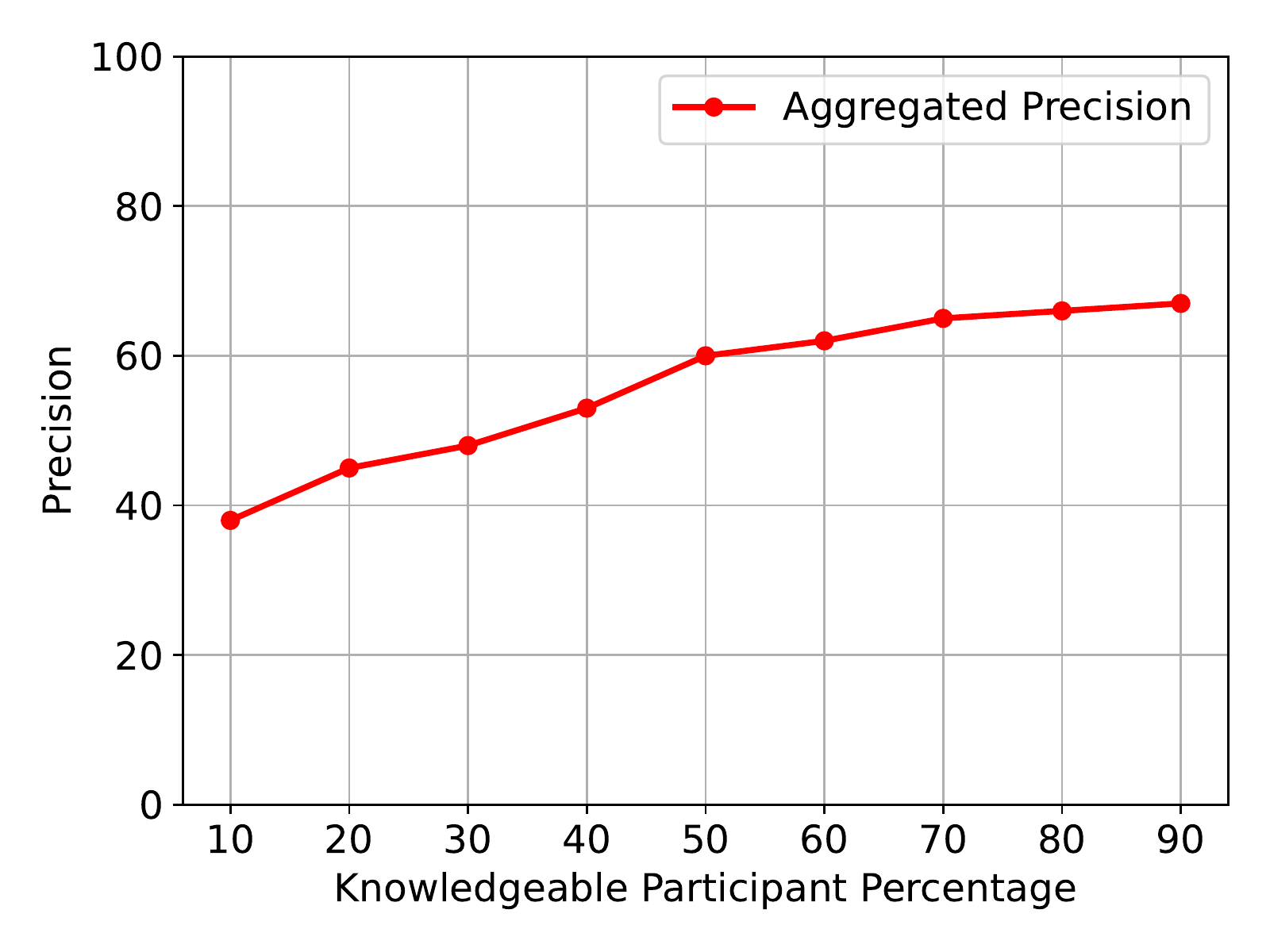}}
 	\caption{Precision of \systemName for varying percentage of knowledgeable participants.} %
	\label{fig:varyknowledgeableparticipants}
\end{figure}

\descr{Varying Knowledgeable Participants.}
We also intend to vary the number of knowledgeable participants to examine their effect on the performance of the aggregated model. 
We vary the ratio of knowledgeable participants from 10\% to 90\% (with 10\% increments).
The experiment is performed for the knowledgeable participant distribution presented in Section~\ref{sec:distributions}.
Fig.~\ref{fig:varyknowledgeableparticipants} depicts the results for different percentages of knowledgeable participants.  %

\subsection{Discussion}
\label{utilitydiscussion}
\descr{Model Performance.} Overall, Table~\ref{tab:utilitymeasurement} shows that it is feasible to train a federated RNN model to predict security events using \systemName.
As discussed before, using FL provides privacy for the participants and removes direct access and collection of security events. 
However, this comes at the cost of a reduction in precision (0.84 to 0.69). 
\systemName performs differently over various distributions. 
Unsurprisingly, the more Non-IID the distribution is, the less utility the aggregated model has.
For instance, in the Extreme Non-IID distribution, the Non-IIDness score is 18.23, while precision and recall are 0.53 and 0.57, respectively.

\descr{Participant's Contribution.} Fig.~\ref{fig:top20influencescorealgo} allows us to reason on how impactful participants contribute to the model. 
We find that the primary distribution does not include a participant with all the classes (what we denote as a knowledgeable participant).
Furthermore, as the baseline aggregated precision is higher in the primary distribution, including the non-impactful participants in the FL training process can improve utility somewhat (from 0.65 to 0.68 precision).
Obviously, if we increase the percentage of knowledgeable participants, the utility would also increase. 

Fig.~\ref{influencetop20extreme} shows that, in the Extreme Non-IID distribution, the knowledge hosted by different participants is highly complementary.
There is no overlapping between participants, so removing any of the participants from the training process will cause a loss of security event information in the whole training dataset. %
This is unlike in the primary or knowledgeable participants distributions, %
as, in these cases, the information about the training data instances hosted by different participants may in fact overlap.
Moreover, we also observe that learning with only a subset of Non-IID participants deteriorates the precision of the model due to more Non-IIDness in the training data and also a smaller number of participants (cf.~Fig.~\ref{influencetop20extreme} vs.~Fig.~\ref{influencetop20primary}). 
However, the deterioration of the aggregated model precision is likely due to the Non-IIDness, as suggested by comparing Fig.~\ref{influencetop20extreme} and Fig.~\ref{influencetop20knowledgeable}. 
The number of participants is similar in both settings. 
Therefore, the precision of the aggregated model is higher when the data distribution is more IID.

\descr{Participant's Benefit.} Fig.~\ref{fig:top20localvsglobal} allows us to assess whether impactful participants benefit from \systemName. 
From Fig.~\ref{top20localvsglobalknowledgeable}, we observe that the majority of impactful participants that are also knowledgeable (i.e., they include instances from all the classes) are not benefiting much from federating, as both curves stay at the same level.
However, Fig.~\ref{top20localvsglobalprimary} shows that there exist impactful participants (around 11 from 20) that do benefit. 
Interestingly, a few organizations even have a deficit in model performance if they participate and use the aggregated model. 
The reason is likely due to the noise prompted in the aggregated model from the non-impactful organizations.
In  Fig.~\ref{top20localvsglobalextreme}, we see that all of the impactful organizations benefit from participating in \systemName in the Extreme Non-IID distribution.

Fig.~\ref{fig:totallocalvsaggregated} shows how all organizations in different distributions benefit from \systemName by separating them into groups of 20\% of entire organizations and averaging the measurements in the groups (the groups are sorted by influence score).
Fig.~\ref{totallocalvsaggregatedprimary} shows that 80\% of the organizations benefit from participating, with an inverse relation to the contribution impact. 
From Fig.~\ref{totallocalvsaggregatedknowledgeable}, we observe that the top 60\% of the organizations with the highest influence score do not benefit, and we know that the percentage of the knowledgeable participants for the distribution is also 60\%. 
This confirms that impactful organizations are knowledgeable ones. 
From Fig.~\ref{totallocalvsaggregatedextreme}, we see that all the organizations benefit from \systemName; also, the amount of benefit is similar in all the groups as each organization includes one type of security event in this Extreme Non-IID distribution. 

Finally, in Fig.~\ref{fig:varyknowledgeableparticipants}, we vary the percentage of knowledgeable participants. 
Above 50\%, the precision of the aggregated model does not increase by a substantial amount, showing that the trained model would not improve further. 

\descr{Communication Overhead.} Since the communication overhead is relatively limited (lower than uploading the raw data) and triggered in an ad-hoc manner when resources (including bandwidth) are available, it is not a significant concern in \systemName. 
Nevertheless, we provide a brief estimation: the number of memory arrays and hidden LSTM memory array units is 4 and 128, respectively (as in~\cite{shen2018tiresias}). 
This yields a local model of size around 30 MB.
Considering that 1) less than 200 rounds of FL are usually enough for the model to converge, 2) the process is not triggered frequently (order of once a day), and 3) we do not envision a deployment involving mobile devices, we believe this constitutes a negligible communication overhead in the enterprise world.
Moreover, compression frameworks like fedzip~\cite{malekijoo2021fedzip} can be used to further reduce communication overhead.

\section{Robustness}\label{sec:robustness}
In this section, we evaluate the robustness of \systemName against data poisoning noise injected by organizations. 
As discussed in Section~\ref{preliminary:attacks}, we experiment with the distributed backdoor attack proposed in \cite{bagdasaryan2020backdoor} (designing new backdoor poisoning methods is beyond the scope of our work). 

Backdoor attacks yield attacker-desired misclassifications only on particular inputs embedded with a pre-defined trigger pattern; otherwise, classification performance remains unaffected~\cite{Gu2019backdoor}. 
Compared to untargeted data poisoning, these attacks are more difficult to detect and mitigate, mostly due to the excess capacity of modern deep neural network-based classifiers \cite{manoj2021excess}. 
During the attack, the adversary injects training instances embedded with the trigger pattern and the attacker-specified class labels. 
These backdoored training instances are then part of the training process and bias the decision output of the model.

In FL, there is no direct way to perform any centralized ``verification'' on participants'  training data, as, besides communication efficiency, the main goal of FL is that training data should not be disclosed, neither to the aggregation server nor to other participants~\cite{bagdasaryan2020backdoor,bhagoji2019analyzing}.
Therefore, backdoor attacks, as an intrinsic and stealthy threat to the integrity of FL-trained models, constitute the focus of our study over the robustness valuation of \systemName.%

\subsection{Experimental Setup}
\label{sec:experimentsetup}
We perform a backdoor attack following these steps: %
\begin{enumerate}[\textbf{Step} 1:]
\item Iterate over all machines in the compromised organization.
\item Iterate over all the series of security events. 
\item For each sequence of security events, if the class is $e_{3642}$, add $e_{0}$ after that.\footnote{We pick $e_{3642}$ as it has the highest frequency before the target event in our dataset.}
If the class is not $e_{3642}$, add $e_{3642}$, $e_{0}$ at the end of the sequence. %
\end{enumerate}
Overall, the goal of the backdoor attack is to make the aggregated model predict class $e_{0}$ after security event $e_{3642}$.
 
\descr{Settings.} We perform the attack on the three distributions. 
The number of attackers is set to 1\% of the total organizations in a round of FL aggregation. 
We set the participation rate to 1 so that the server selects all the participants. 
Moreover, the attack is performed on every epoch of training. 
This corresponds to a rather strong backdoor poisoning attack setting, as we aim to study the utility deterioration of \systemName under the worst-case scenario. 

\descr{Performing the attack on beginning or final FL rounds.} Here, the adversary performs the attack on every round of the FL process.
We also experiment with attacking the first 10 and the last 10 rounds of the process. 
We evaluate the effectiveness of the attack by measuring its accuracy, i.e., the accuracy of the backdoored model on the backdoored data and main task precision.

\subsection{Defenses}

We also apply the defense methods against the backdoor attack in FL, as presented in Section~\ref{preliminary:robustness}. %

For trimmed mean~\cite{yin2018byzantine}, we set $\beta$ to 0.1. 
For norm bounding~\cite{sun2019can}, we select the bound as 5, and in weak DP~\cite{sun2019can}, the injected noise is from Gaussian distribution with variance $\sigma = 0.05$. 
In CDP, we experiment with $\epsilon = 3.8$ and $\delta = 10^{-5}$.

Overall, we measure the effectiveness of the distributed backdoor attack against \systemName by evaluating the prediction accuracy of the main task and backdoor-embedded input instances. 

\begin{table}
\small
\begin{center}
\setlength{\tabcolsep}{2pt}
\begin{tabular}{lrrr}
\toprule
\textbf{Distribution}& \textbf{\#Attackers}& \textbf{Attack}& \textbf{Main Task}  \\  
& {\bf (1\%)} & {\bf Accuracy} & {\bf Precision}\\
\midrule
Primary &37   &0.94   &0.65       \\ 
Knowledgeable Participant & 14 & 0.90 & 0.61      \\ 
Extreme Non-IID& 7 & 0.89 & 0.50  	  \\ 
\bottomrule
\end{tabular}
\end{center}
\caption{Backdoor Attack and Main Task Performance.}
\label{tab:backdoorattack3dist}
\vspace{-0.5cm}
\end{table}

\subsection{Results}\label{robustness:discussion}

Table~\ref{tab:backdoorattack3dist} presents the results of the distributed backdoor attack. 
For all three distributions, the attack is quite effective even with just 1\% of compromised participants, while the prediction precision of the main task is hardly affected.
For instance, in the primary distribution, with 37 compromised organizations, the attack accuracy over the backdoor poisoned testing instances is as high as 0.94, but the main task precision only decreases from 0.69 to 0.65.

In Table~\ref{tab:backdoorfirstlast}, we report the results of the backdoor attack performed on the initial and final rounds. 
Again, the main task precision is unchanged (only in Extreme Non-IID distribution does it decrease from 0.53 to 0.52 when the attack is performed in the last 10 rounds). 
The attack is more effective when performed in the last rounds rather than the initial ones. 
That is because the model is towards convergence in the final rounds, and performing the attack impacts the model more. %

The results of the experiments with the defenses are presented in Table~\ref{tab:defensesrobustness}.
Here we report precision, recall, F1 score, and accuracy of both the main task and the backdoor task, again for all three distributions. 
Overall, Trimmed Mean and Krum are ineffective for extreme Non-IID distribution. 
However, they defend against attacks in the primary and knowledgeable participant distributions. 
This might be because both defense methods assume that the poisoned local model updates should significantly differ from the aggregated global model; in Non-IID distributions, that does not apply. 
Norm bounding, weak DP, and CDP defend better across the board, although at the cost of degrading the performance of the main task.

\begin{table}[t]
\small
\setlength{\tabcolsep}{3pt}
\begin{tabular}{l@{}rrrr}
\toprule
\textbf{Distributions} &
  \multicolumn{2}{c}{\textbf{\begin{tabular}[c]{@{}c@{}}Attack\\ Accuracy\end{tabular}}} &
  \multicolumn{2}{c}{\textbf{\begin{tabular}[c]{@{}c@{}}Main Task\\ Precision\end{tabular}}} \\ \midrule
\textbf{}            & \textbf{First 10} & \textbf{Last 10} & \textbf{First 10} & \textbf{Last 10} \\ \specialrule{0.05pt}{1pt}{1pt}
Primary              & 0.64              & 0.78             & 0.69              & 0.69             \\
Knowledgeable Participant & 0.63              & 0.79             & 0.62              & 0.62             \\
Extreme Non-IID      & 0.59              & 0.75             & 0.53              & 0.52             \\ \bottomrule
\end{tabular}
\caption{Backdoor attack on the first and last 10 rounds.}
\label{tab:backdoorfirstlast}
\vspace{0.2cm}
\end{table}

\begin{table}[t]
\small
\setlength{\tabcolsep}{3pt}
\begin{tabular}{ll|rrrr|rrrr}
\toprule
                         &               & \multicolumn{4}{c|}{\bf Main Task}              & \multicolumn{4}{c}{\bf Attack}                 \\ 
                         &               & \bf prec. & \bf rec.   & \bf F1       & \bf acc. & \bf prec. & \bf rec.   & \bf F1       & \bf acc. \\ \midrule
\multirow{8}{*}{\rotatebox[origin=c]{90}{\textbf{Primary}}} & No Def.    & 0.65      & 0.68     & 0.66     & 0.73     & 0.87      & 0.90      & 0.88     & 0.94     \\
                         & Trimmed M.\hspace{-0.05cm} & 0.61      & 0.59     & 0.59     & 0.65     & 0.63      & 0.65     & 0.63     & 0.68     \\
                         & Krum          & 0.62      & 0.60     & 0.60     & {\bf 0.67} & 0.65      & 0.67     & 0.66     & 0.71     \\
                         & FLTrust       & {\bf 0.63}  & 0.52     & 0.57     & {\bf 0.67} & 0.60      & 0.64     & 0.62     & 0.66     \\
                         & DnC           & {\bf 0.63}  & {\bf 0.62} & {\bf 0.62} & 0.64     & 0.66      & 0.67     & 0.66     & 0.69     \\
                         & Norm B. & 0.60      & 0.55     & 0.57     & 0.63     & 0.58      & 0.62     & 0.60     & 0.57     \\
                         & Weak DP       & 0.56      & 0.53     & 0.54     & 0.59     & 0.54      & 0.51     & 0.52     & 0.54     \\
                         & CDP ($\epsilon$=3.8)          & 0.47      & 0.49     & 0.47     & 0.55     & {\bf 0.33}  & {\bf 0.42} & {\bf 0.37} & {\bf 0.43} \\ \midrule
\multirow{8}{*}{\rotatebox[origin=c]{90}{\textbf{Knowledgeable Part.}}}    & No Def.    & 0.61      & 0.63     & 0.62     & 0.69     & 0.88      & 0.89     & 0.88     & 0.90      \\
                         & Trimmed M.\hspace{-0.05cm}  & 0.59      & 0.57     & 0.58     & 0.67     & 0.68      & 0.70      & 0.69     & 0.71     \\
                         & Krum          & 0.55      & {\bf 0.61} & 0.57     & {\bf 0.68} & 0.69      & 0.74     & 0.71     & 0.68     \\
                         & FLTrust       & 0.57      & 0.60     & 0.58     & 0.67     & 0.69      & 0.70      & 0.69     & 0.65     \\
                         & DnC           & {\bf 0.60}  & 0.59     & {\bf 0.59} & 0.65     & 0.64      & 0.67     & 0.65     & 0.66     \\
                         & Norm B. & 0.52      & 0.57     & 0.54     & 0.63     & 0.55      & 0.63     & 0.59     & 0.63     \\
                         & Weak DP       & 0.50      & 0.51     & 0.50     & 0.57     & 0.52      & 0.56     & 0.54     & 0.60     \\
                         & CDP ($\epsilon$=3.8)           & 0.46      & 0.44     & 0.45     & 0.52     & {\bf 0.43}  & {\bf 0.46} & {\bf 0.44} & {\bf 0.57} \\ \midrule
\multirow{8}{*}{\rotatebox[origin=c]{90}{\textbf{Extreme Non-IID}}} & No Def.    & 0.50      & 0.55     & 0.52     & 0.63     & 0.89      & 0.86     & 0.87     & 0.89     \\
                         & Trimmed M.\hspace{-0.05cm}  & 0.47      & 0.52     & 0.49     & 0.60     & 0.85      & 0.85     & 0.85     & 0.83     \\
                         & Krum          & {\bf 0.48}  & {\bf 0.54} & {\bf 0.51} & {\bf 0.61} & 0.78      & 0.80      & 0.79     & 0.85     \\
                         & FLTrust       & {\bf 0.48}  & 0.50     & 0.49     & {\bf 0.61} & 0.73      & 0.74     & 0.73     & 0.68     \\
                         & DnC           & 0.46      & 0.49     & 0.47     & 0.59     & 0.79      & 0.80      & 0.79     & 0.77     \\
                         & Norm B. & 0.41      & 0.39     & 0.40     & 0.58     & 0.63      & 0.69     & 0.66     & 0.59     \\
                         & Weak DP       & 0.40      & 0.37     & 0.38     & 0.54     & 0.57      & 0.61     & 0.59     & 0.55     \\
                         & CDP ($\epsilon$=3.8)          & 0.36      & 0.34     & 0.35     & 0.48     & {\bf 0.47}  & {\bf 0.52} & {\bf 0.49} & {\bf 0.48}\\ \bottomrule
\end{tabular}
\caption{Evaluation of Robustness Defenses.} 
\label{tab:defensesrobustness}
\vspace{0.2cm}
\end{table}

\section{Privacy}\label{sec:privacy}
Last but not least, we assess the resilience of \systemName to privacy leakage---specifically, performing the two membership inference attacks presented in Section~\ref{preliminary:attacks}.

We focus on membership inference because, when a record is known to the adversary, learning that it was used to train a particular model indicates information leakage through the model. 
Overall, these kinds of attacks are often considered a ``measuring stick'' that access to a model leads to potentially serious privacy leakage, and in fact they are often gateways to further attacks~\cite{de2021critical}.

\subsection{Experimental Setup}
Membership inference against FL has only been done successfully on settings involving a small number of participants~\cite{melis2019exploiting,nasr2019comprehensive,de2021critical,zhang2020gan}.
This is due to the signal of any participant's input naturally weakening with an increasing number of participants.

Therefore, we need to decrease the number of participants to drive any meaningful experimental results.
To this end, we move away from the synthesized distribution settings discussed in Section~\ref{sec:distributions} and pool all the security events and redistribute them randomly.
For Nasr et al.'s attack~\cite{nasr2019comprehensive}, we experiment with 2, 3, 4, and 5 organizations participating in the federated model aggregation.
For Zhang et al.'s attack~\cite{zhang2020gan}, we distribute the pooled dataset among 2 and 3 organizations.
(The attacks do not work beyond this number of participants.)
In both attacks, one of the organizations is the adversary performing the membership inference attack to infer whether or not specific security events datasets are included in other organizations' training data.

\begin{table}[t]
\small
\setlength{\tabcolsep}{3pt}

\begin{tabular}{rcrr}
\toprule
\textbf{\#Organizations}                                  && \textbf{Nasr et al.~\cite{nasr2019comprehensive}}   & \textbf{Zhang et al.~\cite{zhang2020gan}}   \\ \midrule
2                                && {0.78}       &   {0.75}     \\ 
3                                && {0.72}       &   {0.69}     \\ 
4                                && {0.58}       &    {--}    \\ 
5                                && {0.54}       &   {--}     \\ \bottomrule
\end{tabular}
\caption{Membership Inference Attack Accuracy.}
\label{tab:miattack}
\vspace{0.2cm}
\end{table}

\subsection{Results}\label{subsec:privacy}
Table~\ref{tab:miattack} reports the accuracy of the attacks for an increasing number of participants.
Note that the baseline for membership inference attack is a random guess with 50\% accuracy (a data record {\em is} or {\em is not} part of the training set).

Overall, Nasr et al.'s attack~\cite{nasr2019comprehensive} is more effective than Zhang et al.'s ~\cite{zhang2020gan}.
However, neither attacks are successful when more than a handful of organizations participate (i.e., accuracy quickly reaches random guess baseline).
Therefore, \systemName should not generally be exposed to privacy leakage attacks as it usually consists of many organizations.
However, it does indicate that one should be very careful and seriously consider privacy risks when involving a limited number of participants.

\descr{Defenses.} Overall, Centralized Differential Privacy (CDP, also known as participant-level DP) can be used to reduce the accuracy of membership inference attacks. 
However, in our experiments, we find that applying CDP prevents the aggregated model from converging. 
This is likely due to two reasons.
First, as the number of participants is small, the noise reduces the stability of the model training process, which eventually causes the divergences of the model aggregation. 
Second, %
the amount of noise needed to be added is relatively large, which severely affects the performance of the model.
because of the small number of participants and the complexity of the model. 
As a result, adopting CDP as a mitigation strategy to prevent privacy leakage with a small number of participants is likely ineffective, as also found in prior work~\cite{melis2019exploiting,DBLP:journals/corr/abs-2009-03561}.

\section{Related Work}
This section reviews previous work on predicting security events, applications of FL to security, as well as measurements of utility, robustness, and privacy in FL.

\subsection{Prediction of Security Events}

\descr{Forecasting Security Postures.} 
Prior work has used machine learning to forecast security postures.
The main intuition is to learn how to do so by training a model using historical data (i.e., metadata profiling previous security postures or historical security events collected between $t_0$ and $t_i$).
At timestamp $t_{k+1}$, the model produce a binary prediction outcome (i.e., if a breach or an attack is likely to happen) using present data (i.e., data collected between $t_{i+1}$ and $t_k$)~\cite{bilge2017riskteller,Liu2015usenix,Soska2014usenix,sharif2018predicting,Sabottke2015usenix}. 

In~\cite{Liu2015usenix}, multiple features are defined to describe mismanagement symptoms (e.g., misconfigured DNS) and malicious activities (e.g., scanning activities originating from this organization's network) of an organization's network. 
A random forest classifier is then used to forecast security incidents.
Soska et al.~\cite{Soska2014usenix} characterize websites using network traffic statistics, webpage structures, and contents; the profiling features are then fed into a C4.5 decision tree classifier to predict whether a given website will become malicious in the future. 
Possible vulnerability exploits have also been predicted using information discussed on Twitter, including Twitter messages and Common Vulnerability Scoring System (CVSS) information~\cite{Sabottke2015usenix}.

\descr{Deep Learning (DL) based Approaches.} 
In recent years, DL-based security event prediction methods~\cite{deeplog,deepcase,shen2018tiresias} have been employed to predict the actions that will be taken by an attacker. 
Typically, these methods capture the sequential profiles of security event logs of normal system sessions using DL-based time series models, such as Recurrent Neural Nets (RNN) \cite{deeplog}, Long Short-Term Memory (LSTM)~\cite{deepcase}, and Gated Recurrent Unit (GRU)~\cite{shen2018tiresias}. 

Specifically, given the first $K$ log entries $\{e_{t-K},...,e_{t-2},e_{t-1}\}$ as input, the time series model is trained to predict the successive log $e_{t}$.
Based on the log prediction results, these methods can flag the log sequences that are deviated significantly from the normal system execution traces as anomaly incidents. 
For instance, \textit{DeepCase} core prediction model is a Recurrent Neural Network (RNN) enhanced with a self-attention mechanism. 
The attention mechanism weights of the derived RNN model indicate the relevance between each input historical log entry and the target log entry to predict.

Treating the integer log indexes as class labels, both \textit{DeepLog}~\cite{deeplog} and \textit{DeepCase}~\cite{deepcase} conduct log entry prediction as a problem of multi-class classification. 
They adopt the top-$K$ prediction scheme: they check if the target log entry is one of the top $K$ predictions (the $K$ predicted log entries with the highest classification confidence). %

Shen et al.~\cite{shen2018tiresias} develop a system called Tiresias for predicting security events through deep learning that leverages recurrent neural networks to predict future events on a machine based on previous observations.
The authors test Tiresias on a dataset of 3.4 billion security events collected from a commercial intrusion prevention system; Tiresias is effective in predicting the next event that will occur on a machine with a precision of up to 0.93.

Finally, general-purpose tools like Log2Vec~\cite{liu2019log2vec}, Attack2Vec~\cite{shen2019attack2vec}, and ATLAS~\cite{alsaheel2021atlas} have been presented that apply natural language processing to cybersecurity areas.
As opposed to this line of our work, which entails a centralized collection of security events, we use a FL-based approach. %

\subsection{Federated Learning (FL) for Security}
While \systemName uses FL to predict security events, FL has also been used in security applications, ranging from intrusion detection to anomaly detection, etc.
For instance, Li et al.~\cite{li2020deepfed} present the DeepFed framework to collaboratively build intrusion detection models in industrial cyber-physical systems.
Chen et al.~\cite{chen2020intrusion} present the FL-based Attention Gated Recurrent Unit (FedAGRU), an intrusion detection algorithm for wireless edge networks which prevents uploading parameters that do not benefit the overall model, thus decreasing communication overhead.

Kang et al.~\cite{kang2019incentive} study worker selection and incentive mechanism issues for reliable FL in mobile networks. 
Liu et al.~\cite{liu2020deep} propose an FL-based deep anomaly detection framework for sensing time-series data in industrial products in the Internet of Things; the model uses attention mechanism-based CNNs to capture important fine-grained features and prevent memory loss and gradient dispersion problems.
Finally, G\'alvez et al.~\cite{galvez2020less} use FL for Android malware detection and study the effect of poisoning and membership inference attacks against the framework.

\subsection{Utility, Robustness, and Privacy Measurements in FL}
In the previous sections, we have already reviewed some results focusing on utility, robustness, and privacy in FL.
In addition, Wang et al.~\cite{wang2019measure} propose group instance deletion and Shapley values to calculate participant contribution in FL, aiming to support meaningful credit and reward allocations.  
Also, Song et al.~\cite{song2019profit} measure the contribution of participants in horizontal FL by defining the contribution index based on the Shapely value. 
Yu et al.~\cite{yu2020salvaging} study how local adaption techniques help improve the utility of private FL models for participants.  

Prior work has looked at byzantine attacks in FL, which compromise the global model via arbitrarily malicious gradients or intentionally crafted local model updates~\cite{damaskinos2019aggregathor,blanchard2017machine,Shejwalkar2021ManipulatingTB,Cao2021FLTrustBF}.
For instance, backdoor attacks presented in Section~\ref{preliminary:attacks} make the global model output the target label specified by the adversary for particular examples~\cite{bagdasaryan2020backdoor,fung2020limitations,bhagoji2019analyzing,sun2019can,Wang2020nips}.
In this context, Sattler et al.~\cite{sattler2020byzantine} analyze the use of clustered FL where participants are grouped based on similarities between their parameter updates to provide robustness. 

Finally, previous work has quantified information leakage from exchanging gradients in FL~\cite{melis2019exploiting,zhu2019deep,nasr2019comprehensive}.
Moreover, Jourdan et al.~\cite{jourdan2021privacy} study utility-vs-privacy trade-offs in FL using private personalized layers and experiment with membership and property inference attacks. 
They find that personalized layers speed up the model's convergence and better mitigate inference attacks.

\section{Discussion \& Conclusion}
\label{sec:discussionconclusion}

\descr{Recap.} In this paper, we experimented with using Federated Learning (FL) for collaboratively training machine learning models and predicting security events.
More precisely, we evaluated the model performance compared to a centralized approach, where all security events from all organizations are pooled at a central server; then, we analyzed the robustness of the federated model to distributed backdoor poisoning attacks and privacy leakage through membership inference attacks. 
In the process, we introduced \systemName, a system using FL to train a Recurrent Neural Network (RNN) for predicting security events in a privacy-friendly, distributed way.
We trained \systemName over a dataset obtained from a major security company (involving over 34 million security events and 2 million machines) and conducted several experiments over several different data distributions, aiming to simulate different levels of heterogeneity and their effect on collaborative learning.

\descr{Model Performance.} We find that model performance degrades, although slightly, in the federated setting compared to a centralized approach. 
However, we believe this could be a reasonable price as it enables settings that would not otherwise be possible, as sensitive security events often cannot be shared across different organizations.
Moreover, ours is only the first attempt at the problem, and other collaborative learning techniques like~\cite{cutkosky2018distributed,meng2016asynchronous,smith2018cocoa} could be explored that have the potential of working better than FL.

Overall, we show that certain data distribution settings may be significantly more or less ``suitable'' to FL.
For instance, the primary distribution, which follows a realistic distribution of data in the real world, or the knowledgeable participant distribution, which consists of participants with rich datasets, are appropriate distributions to be used in \systemName, as the final model utility is acceptable and the participants can benefit from it.
However, in highly non-IID distributions, the final model utility is not high. 
Although, the individuals would benefit as the utility can still be higher than theirs.
Hence, our work suggests that the FL-based approach for security event prediction is a viable approach, for the time being, only when data is distributed in a certain way across organizations.

\descr{False Positives.} As discussed in Section~\ref{modelperformance}, \systemName yields non-negligible false positive rates.
However, we stress that our focus is on predicting the type of the possible incidents that are likely to occur in the future based on historical event observations rather than detecting/categorizing the anomalies/infection that already occurred (as in most malware detection settings). 
Thus, performance is better evaluated using precision rather than FPR, as done in previous work~\cite{shen2018tiresias,deepcase}.
In fact, relatively high FPRs occur in prior work as well~\cite{liu2015cloudy,shen2018tiresias,vance2019fog}.
Furthermore, this kind of system predicts events; our RNN model is an encoding of the sequential pattern of attack events. 
This means that, rather than solely predicting future incidents and taking immediate blocking/defense actions, analysts used them to understand how attack events are chained together and shed light on the sequential patterns of incidents and, thus, on the relations between security events.

Nonetheless, tuning FPRs and overall responding to alerts remains an open challenge, both in general (e.g., with security incident and event management tools or intrusion detection systems) and specifically for our line of work.
Ongoing research has been studying the impact of security alerts and discussed ways to improve how security warnings can be effectively delivered to SecOps, e.g.,~\cite{anderson2015polymorphic,petelka2019put,MSalert}; integrating these techniques with \systemName is an interesting item for future work.

\descr{Future Work.} We plan to experiment with different FL instantiations and datasets, aiming to improve model accuracy and assess the generalizability of our results. 
We are confident that follow-up work can experiment with different instantiations of collaborative/federated learning, as well as improve Differential Privacy bounds and its variants against robustness and privacy attacks.

\descr{Acknowledgments.} This work was (partially) funded by a Microsoft EPSRC Case Studentship, an Amazon Research Award, and the National Science Foundation under grant CNS-2127232.
\bibliographystyle{ACM-Reference-Format}
%\bibliography{ref}
%%% -*-BibTeX-*-
%%% Do NOT edit. File created by BibTeX with style
%%% ACM-Reference-Format-Journals [18-Jan-2012].

\end{document}